\documentclass{emulateapj}

\usepackage{amsmath}
\usepackage{gensymb}
\usepackage{upgreek}

\def\aj{AJ}%
\def\apj{ApJ}%
\def\apjl{ApJ}%
\def\apjs{ApJS}%
\def\ao{Appl.~Opt.}%
\def\aap{A\&A}%
\def\mnras{MNRAS}%
\def\pasp{PASP}%
\def\nat{Nature}%

\def\mean#1{\left< #1 \right>}

\renewcommand{\thefootnote}{\fnsymbol{footnote}}

\slugcomment{Revised version 2}

\shortauthors{Defr\`ere et al.}
\begin{document}

\title{Nulling Data Reduction and On-Sky Performance of the Large Binocular\\ Telescope Interferometer}
\author{D.~Defr\`ere\altaffilmark{1}, P.M.~Hinz\altaffilmark{1}, B.~Mennesson\altaffilmark{2}, W.F.~Hoffmann\altaffilmark{1}, R.~Millan-Gabet\altaffilmark{3}, A.J.~Skemer\altaffilmark{1}, V.~Bailey\altaffilmark{1,4}, W.C.~Danchi\altaffilmark{5}, E.C.~Downey\altaffilmark{1},  O.~Durney\altaffilmark{1}, P.~Grenz\altaffilmark{1}, J.M.~Hill\altaffilmark{6}, T.J.~McMahon\altaffilmark{1}, M.~Montoya\altaffilmark{1}, E.~Spalding\altaffilmark{1}, A.~Vaz\altaffilmark{1}, O.~Absil\altaffilmark{7}, P.~Arbo\altaffilmark{1}, H.~Bailey\altaffilmark{8}, G.~Brusa\altaffilmark{1}, G.~Bryden\altaffilmark{2}, S.~Esposito\altaffilmark{9}, A.~Gaspar\altaffilmark{1}, C.A.~Haniff\altaffilmark{10}, G.M.~Kennedy\altaffilmark{11}, J.M.~Leisenring\altaffilmark{1}, L.~Marion\altaffilmark{7}, M.~Nowak\altaffilmark{3,12}, E.~Pinna\altaffilmark{9}, K.~Powell\altaffilmark{1}, A.~Puglisi\altaffilmark{9}, G.~Rieke\altaffilmark{1}, A.~Roberge\altaffilmark{5}, E.~Serabyn\altaffilmark{2}, R.~Sosa\altaffilmark{1}, K.~Stapeldfeldt\altaffilmark{5}, K.~Su\altaffilmark{1}, A.J.~Weinberger\altaffilmark{12}, and M.C.~Wyatt\altaffilmark{11}}

\affil{\altaffilmark{1}Steward Observatory, Department of Astronomy, University of Arizona, 933 N. Cherry Ave, Tucson, AZ 85721, USA}
\affil{\altaffilmark{2}Jet Propulsion Laboratory, California Institute of Technology, 4800 Oak Grove Drive, Pasadena CA 91109-8099, USA}
\affil{\altaffilmark{3}NASA Exoplanet Science Institute, California Institute of Technology, 770 South Wilson Avenue, Pasadena CA 91125, USA}
\affil{\altaffilmark{4}Kavli Institute for Particle Astrophysics and Cosmology, Stanford University, Stanford, CA 94305, USA}
\affil{\altaffilmark{5}NASA Goddard Space Flight Center, Exoplanets \& Stellar Astrophysics Laboratory, Code 667, Greenbelt, MD 20771, USA}
\affil{\altaffilmark{6}Large Binocular Telescope Observatory, University of Arizona, 933 N. Cherry Ave, Tucson, AZ 85721, USA}
\affil{\altaffilmark{7}Institut d'Astrophysique et de G\'eophysique, Universit\'e de Li\`ege, 19c All\'ee du Six Ao\^ut, B-4000 Sart Tilman, Belgium}
\affil{\altaffilmark{8}Lunar and Planetary Laboratory, University of Arizona, 1541 E, University Blvd, Tucson, AZ 85721, USA}
\affil{\altaffilmark{9}INAF-Osservatorio Astrofisico di Arcetri, Largo E. Fermi 5, I-50125 Firenze, Italy}
\affil{\altaffilmark{10}Cavendish Laboratory, University of Cambridge, JJ Thomson Avenue, Cambridge CB3 0HE, UK}
\affil{\altaffilmark{11}Institute of Astronomy, University of Cambridge, Madingley Road, Cambridge CB3 0HA, UK}
\affil{\altaffilmark{12}LESIA/Observatoire de Paris, CNRS, UPMC, Universit\'e Paris Diderot, 5 place Jules Janssen, 92195 Meudon, France}
\affil{\altaffilmark{13}Department of Terrestrial Magnetism, Carnegie Institution of Washington, 5241 Broad Branch Road NW, Washington, DC, 20015, USA}

\email{ddefrere@email.arizona.edu}

\begin{abstract}
The Large Binocular Telescope Interferometer (LBTI) is a versatile instrument designed for high-angular resolution and high-contrast infrared imaging (1.5-13\,$\upmu$m). In this paper, we focus on the mid-infrared (8-13\,$\upmu$m) nulling mode and present its theory of operation, data reduction, and on-sky performance as of the end of the commissioning phase in March 2015. With an interferometric baseline of 14.4 meters, the LBTI nuller is specifically tuned to resolve the habitable zone of nearby main-sequence stars, where warm exozodiacal dust emission peaks. Measuring the exozodi luminosity function of nearby main-sequence stars is a key milestone to prepare for future exoEarth direct imaging instruments. Thanks to recent progress in wavefront control and phase stabilization, as well as in data reduction techniques, the LBTI demonstrated in February 2015 a calibrated null accuracy of 0.05\% over a three-hour long observing sequence on the bright nearby A3V star $\beta$ Leo. This is equivalent to an exozodiacal disk density of 15 to 30~zodi for a Sun-like star located at 10\,pc, depending on the adopted disk model. This result sets a new record for high-contrast mid-infrared interferometric imaging and opens a new window on the study of planetary systems. 
\end{abstract}
\section{Introduction}
\label{sec:intro}

\begin{figure*}[!t]
	\begin{center}
		\includegraphics[width=17.0 cm]{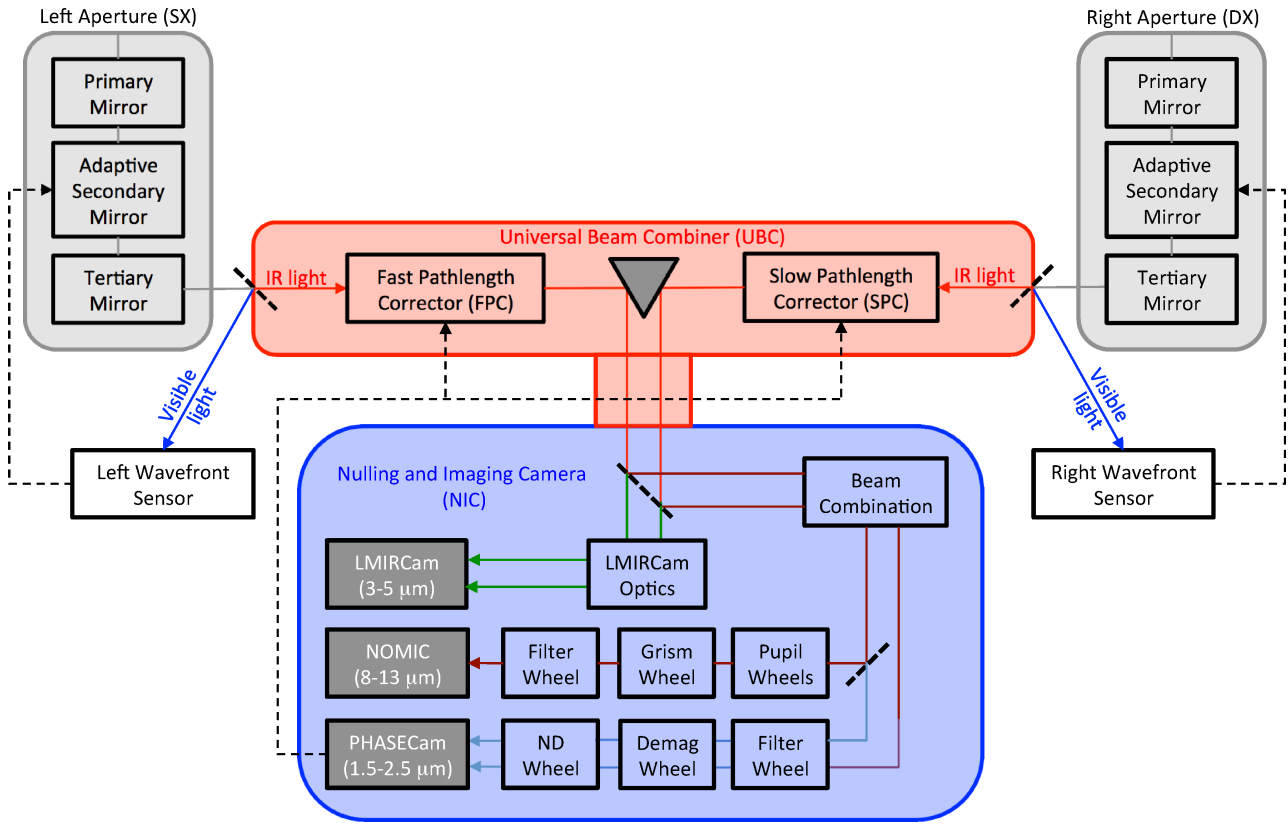}
		\caption{System-level block diagram of LBTI architecture in nulling mode showing the optical path through the telescope, beam combiner (red box), and the NIC cryostat (blue box). After being reflected on LBT primaries, secondaries, and tertiaries, the visible light is reflected on the entrance window and used for wavefront sensing while the infrared light is transmitted into LBTI, where all subsequent optics are cryogenic. The beam combiner directs the light with steerable mirrors and can adjust pathlength for interferometry. Inside the NIC cryostat, the thermal near-infrared (3-5 $\upmu$m) light is directed to LMIRCam for exoplanet imaging, the near-infrared (1.5-2.5 $\upmu$m) light is directed to the phase sensor, which measures the differential tip/tilt and phase between the two primary mirrors, and the mid-infrared (8-13 $\upmu$m) light is directed to NOMIC for nulling interferometry. Both outputs of the beam combiner are directed to the phase and tip/tilt sensor, while only the nulled output of the interferometer is reflected to the NOMIC camera with a short-pass dichroic. The various cameras are shown in dark grey and feed-back signals driving the deformable secondary mirrors and tip-tilt/OPD correctors are represented by dashed lines. Note that this diagram is schematic only and does not show several additional optics.}\label{fig:diagram}
	\end{center}
\end{figure*}

The Large Binocular Telescope Interferometer (LBTI) is an interferometric instrument designed to coherently combine the beams from the two 8.4-m primary mirrors of the LBT for high-angular resolution imaging at infrared wavelengths (1.5-13\,$\upmu$m). It leverages the high sensitivity enabled by LBT's two large apertures, high-quality wavefronts delivered by its adaptive optics systems, low thermal background due to the adaptive secondary architecture, and high angular resolution enabled by the coherent combination of light from the two LBT primary mirrors (14.4\,m center-to-center separation, 22.65\,m maximum baseline). The primary science goal of the LBTI is to determine the prevalence of exozodiacal dust around nearby main-sequence stars in support of a future space telescope aimed at direct imaging and spectroscopy of terrestrial planets (exoEarths) around nearby stars. This warm circumstellar dust, analogous to the interplanetary dust found in the vicinity of the Earth in the Solar system, is produced in comet break-ups and asteroid collisions. Emission and/or scattered light from this dust will be the major source of astrophysical noise for a future visible coronagraph \citep[e.g.,][]{Roberge:2012,Stark:2014} or mid-infrared interferometer \citep[e.g.,][]{Beichman:2006b,Defrere:2010}. As recently discussed by \cite{Stark:2015}, the minimum aperture size for a  future exoEarth detection coronagraphic mission is dependent on several assumed astrophysical quantities among which the most important are the assumed size and optical properties for every Earth-sized planet residing in the habitable zone (HZ), the number of HZ Earth-sized planets per star ($\eta_\oplus$), and the exozodiacal dust cloud surface brightness. While Kepler is currently constraining the two former \cite[e.g.,][]{Burke:2015}, the prevalence of exozodiacal dust in the terrestrial planet region of nearby planetary systems is currently poorly constrained. The bright end of the exozodi luminosity function, i.e.\ several hundred to several thousand times the dust density of the solar zodiacal cloud, has been measured by space-based single-dish telescopes \citep[e.g.,][]{Kennedy:2013} but such a sensitivity is orders of magnitudes too poor to efficiently prepare future exoEarth imaging missions. 

In order to measure fainter exozodiacal disks, a survey of nearby main sequence stars has been carried out with the Keck Interferometer Nuller (KIN). Science results from the KIN were reported recently \citep{Millan-gabet:2011,Mennesson:2014} and indicate that the median level of exozodiacal dust around such stars is no more than 60 times the solar value with high confidence (95\%, assuming a log-normal luminosity distribution). Yet, the state-of-the-art exozodi sensitivity achieved per object by the KIN is approximately one order of magnitude larger than that required to prepare future exoEarth imaging instruments. The LBTI is designed to reach the required level. The first observations, based on commissioning data, were reported recently and showed a sensitivity similar to that of the KIN \citep{Defrere:2015}. These observations were obtained using a coarse fringe tracking algorithm (equivalent to group delay tracking), which is limited to a closed-loop optical path difference (OPD) residual of approximately 1\,$\upmu$m RMS. Phase tracking was commissioned later that year and significantly improved the OPD stability of the system ($\sim$400\,nm RMS). During the same period, improvements in the telemetry tracking also allowed us to test and validate a much more powerful nulling data reduction technique than that used for the first study. This technique, called Nulling Self Calibration (NSC) and pioneered on the Palomar Fiber Nuller \citep[PFN,][]{Hanot:2011, Mennesson:2011}, was adapted for the LBTI and showed to significantly reduce the impact of important systematic errors. Following these results, the LBTI achieved a calibrated null accuracy that is sufficient to start the exozodi survey, called the Hunt for Observable Signature of Terrestrial Systems (HOSTS). The survey will start with a one year science validation phase that will also include some engineering tasks to further improve the performance of the instrument. Overall (including both observations obtained during the commissioning and science validation phases), the HOSTS survey will be carried out over the next two to three years on a sample of 35 to 40 carefully chosen nearby main-sequence stars \citep{Weinberger:2015}. 

This paper provides a description of LBTI's instrumental setup in nulling mode, data reduction, data calibration, and on-sky performance in support of the HOSTS science survey. Section~\ref{sec:architecture} describes the overall architecture of the LBTI with a particular focus on the nulling mode and its on-sky response.  The method used for fringe and tip/tilt tracking is also discussed. Section~\ref{sec:reduction} discusses the level 0 (L0) data products and some observational details including the observing sequence, which was specifically designed for the NSC. Then, the core of this section is dedicated to the level 1 (L1) data reduction process, which consists in converting a set of raw images of various types to raw null measurements. Section~\ref{sec:calib} is dedicated to the null calibration, or level 2 (L2) data reduction, which basically consists in removing the contribution of the instrument from the raw null depth measurements in order to be left with only the contribution from the astrophysical object (or source null). Finally, Section~\ref{sec:perfo} presents on-sky performance of the system at the end of the commissioning phase. This includes throughput, photometric sensitivity, and OPD stability, which are the most relevant metrics to reach high contrasts. Appendices~\ref{app:calib}, \ref{app:trans}, and \ref{app:bckg} provide additional information on the choice of calibrator stars, the spectral transmission of the instrument, and the impact of the background region used for aperture photometry.


\section{Optical setup and methodology} \label{sec:architecture}
\subsection{Overall architecture}

The LBTI is located at the bent center Gregorian focal station of the Large Binocular Telescope \citep[LBT,][]{Hill:2014,Veillet:2014}. The LBT is located on Mount Graham in southeastern Arizona and is operated by an international collaboration among institutions in the United States, Italy, and Germany. It consists of two 8.4-m aperture optical telescopes installed on a single steerable altitude-azimuth mount. This design provides an ideal platform for interferometric observations since it does not require long delay lines and contains relatively few warm optical elements. Both apertures are equipped with deformable secondary mirrors, which are driven with the LBT's adaptive optics system to correct atmospheric turbulence \citep{Esposito:2010, Bailey:2014}. Each deformable mirror uses 672 actuators that routinely correct 500 Zernike modes and provide Strehl ratios as high as 80\%, 95\%, and 99\% at 1.6\,$\upmu$m, 3.8\,$\upmu$m, and 10\,$\upmu$m, respectively \citep{Esposito:2012,Skemer:2014}. 

The overall LBTI system architecture is based on the heritage of the Bracewell Infrared Nulling Cryostat on the MMT \cite[BLINC,][]{Hinz:2000} and will be described in a forthcoming paper (Hinz et al.\ in prep). In this paper, we focus on the parts relevant for nulling interferometry. The LBTI system architecture in nulling mode is represented by the block diagram in Figure~\ref{fig:diagram}. Starlight bounces off the LBT primaries, secondaries, and tertiaries on either side of this figure before coming into the LBTI. Visible light reflects off the LBTI entrance windows and into the adaptive optics wavefront sensors, which control the deformable secondary mirrors. The infrared light transmits into LBTI's universal beam combiner (UBC, see red box) in which all optics are cryogenic. The UBC can direct the light with steerable mirrors and provides a combined focal plane from the two LBT apertures. Beam alignment is done via the Fast Pathlength Corrector (FPC) located in the left part of the UBC and the Slow Pathlength Corrector (SPC) located in the right side. Both the FPC and the SPC can adjust pathlength for interferometry. The FPC provides a Piezo-electric transducer (PZT) fast pathlength correction with 80\,$\upmu$m of physical stroke, capable of introducing 160\,$\upmu$m of optical path difference (OPD) correction. The right mirror provides a larger stroke (40 mm of motion) for slow pathlength correction. In practice, the SPC is used to acquire the fringes while the FPC is used to correct for pathlength variations at high speed (up to 1\,kHz, depending on the magnitude of the star). The SPC can also be used in closed-loop to offload the FPC when it reaches the end of its range. 

Downstream of the UBC, the infrared light enters the cryogenic Nulling and Imaging Camera (NIC), which is equipped with two scientific cameras, i.e.\ LMIRCam \cite[the L and M Infrared Camera,][]{Wilson:2008,Leisenring:2012} and NOMIC \citep[Nulling Optimized Mid-Infrared Camera,][]{Hoffmann:2014}, and a near-infrared fast-readout PICNIC detector (PHASECam) to measure the tip/tilt and phase variations between the LBT apertures. At the entrance of NIC, a trichroic transmits the thermal near-infrared light (3.0-5.0\,$\upmu$m) to the LMIRCam channel and reflects the near-infrared (1.5-2.5\,$\upmu$m) and mid-infrared light (8-13\,$\upmu$m) to the NOMIC channel \citep[see transmission and reflection curve in][]{Skemer:2014}. To minimize non-common path errors, we split the near-infrared and mid-infrared light after beam combination. Both interferometric outputs are directed to the phase and tip/tilt sensor, while only the nulled output of the interferometer is reflected to the NOMIC camera with a short-pass dichroic (see Figure~\ref{fig:combination}). The other output is discarded. In the NOMIC channel, a series of wheels are available to select the wavelength \citep[see list of available filters in][]{Defrere:2015b}, cold stops (pupil wheel), and grisms for low-resolution spectroscopy. In the PHASECam channel, there are several wheels to select the wavelengths (e.g., standard H and K filters), the pupil size, and neutral densities. A more detailed description about these modes can be found in \cite{Hinz:2008}.

\subsection{Nulling mode}

\begin{figure}[!t]
	\begin{center}
		\includegraphics[width=8.5 cm]{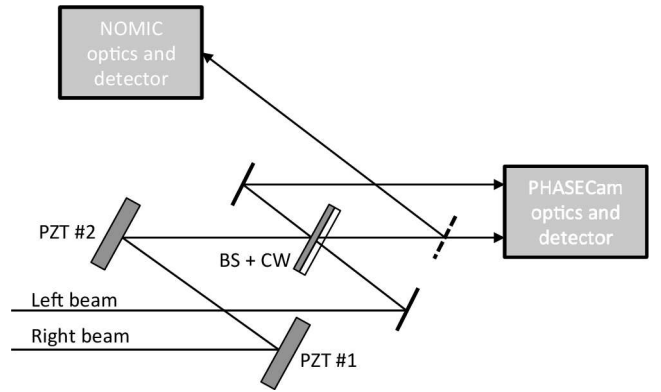}
		\caption{Conceptual schematic of the nulling and PHASECam beam combination. Beam combination is done in the pupil plane on a 50/50 beamsplitter (BS), which can be translated to equalize the pathlengths between the two sides of the interferometer. To achieve an achromatic suppression of light over a sufficiently large bandwidth (8-13\,$\upmu$m), a compensator window (CW) with a suitable thickness of dielectric is introduced in one beam. Both outputs of the interferometer are directed to the near-infrared phase sensor (PHASECam) while one output is reflected to the NOMIC science detector with a short-pass dichroic. Note that this sketch does not show several fold mirrors and biconics. The complete diagram can be found in \cite{Hinz:2008}.}\label{fig:combination}
	\end{center}
\end{figure}

\begin{figure}[!t]
	\begin{center}
		\includegraphics[width=4.2 cm]{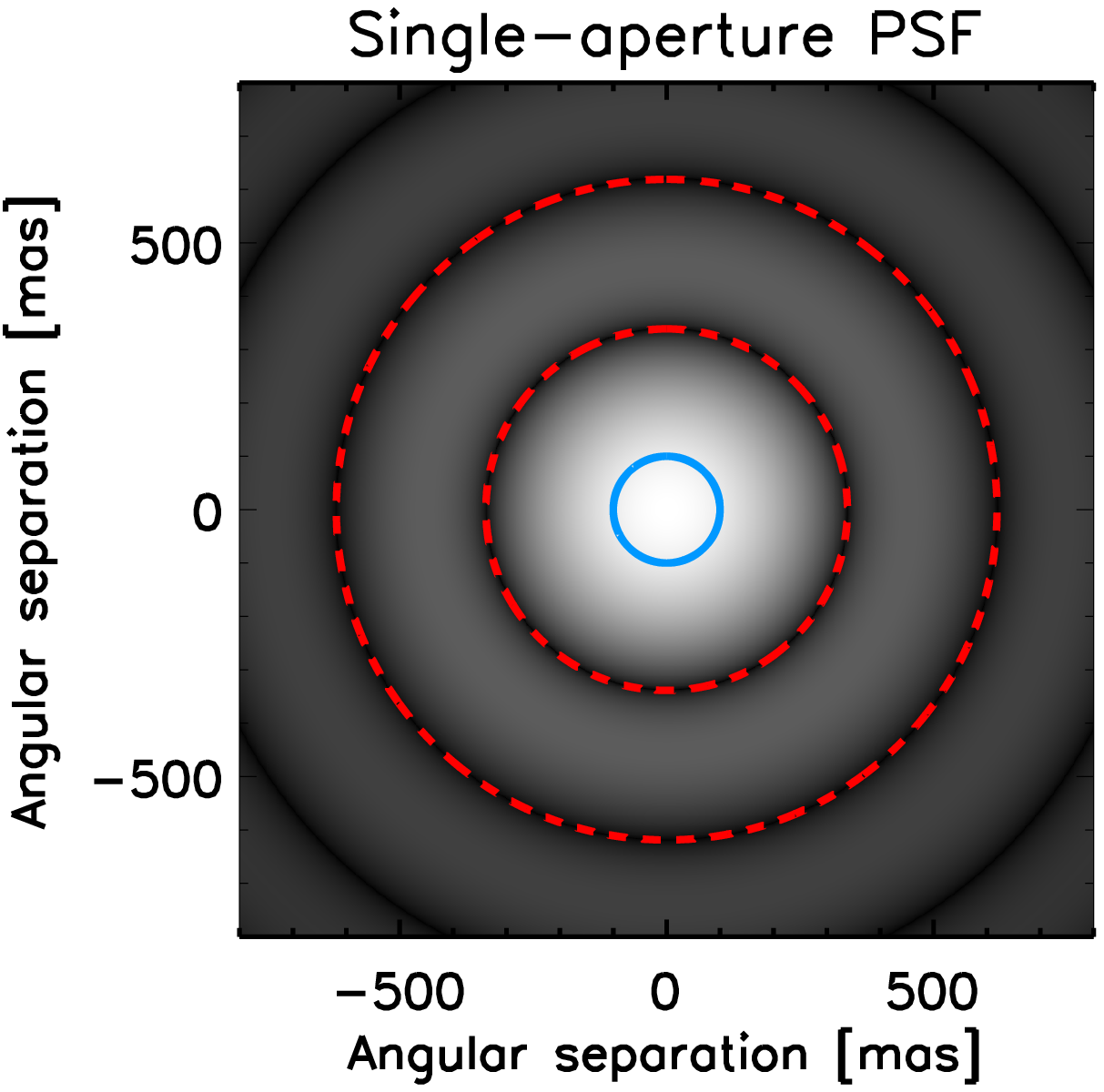}
		\includegraphics[width=4.2 cm]{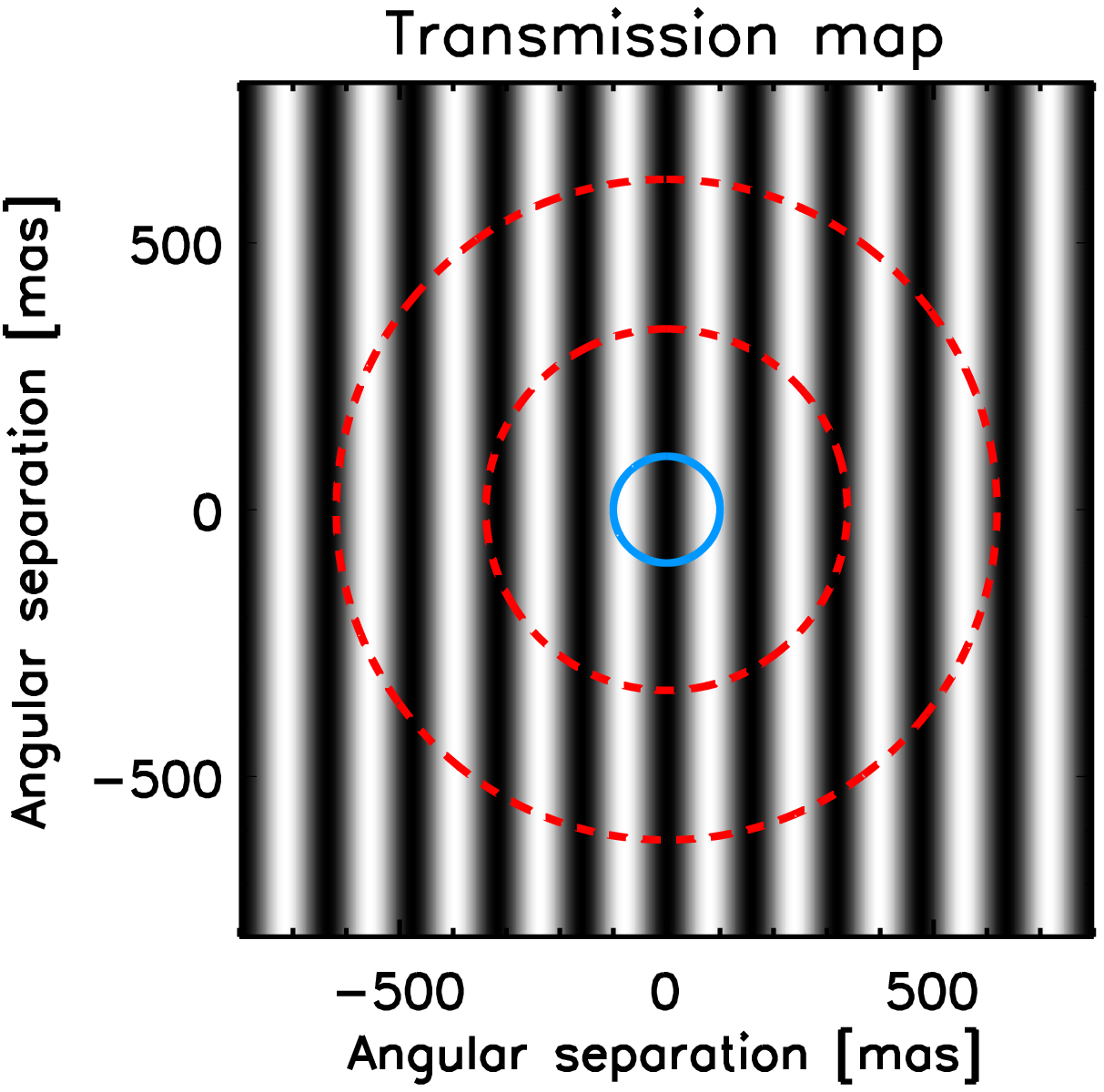}
		\caption{Illustration of LBTI monochromatic single-aperture PSF (left) and interferometric transmission map (right) computed for a wavelength of 11.1\,$\upmu$m over a 1\farcs{5} x 1\farcs{5} field of view, assuming a purely east-west 14.4\,m baseline (north is up, east is to the left). The dashed red lines indicate the position of the first two minima of the single-aperture PSF. The solid blue line indicates the position of Earth's orbit around a sun located at 10\,pc and seen face-on. It is resolved by the interferometer but not by the single-aperture PSF. The PSF is displayed with a square root stretch to better show the first Airy ring while the transmission map is shown with a linear stretch.}\label{fig:tm}
	\end{center}
\end{figure}

One of the main limitations to detect faint circumstellar emission with an infrared interferometer resides in the high dynamic range that must be achieved (e.g., $\sim$10000:1 for LBTI's HOSTS survey). A possible avenue to tackle this observing challenge is to use the undulatory nature of light to perform a destructive interference of the starlight. The technique was first proposed by \cite{Bracewell:1978} to image extra-solar planets and has since then been implemented at various telescopes such as the MMT \citep[e.g.,][]{Hinz:1998c}, the Keck observatory \citep[e.g.,][]{Colavita:2009}, and Hale telescope at Mount Palomar \citep[e.g.,][]{Mennesson:2011}. The basic principle is to combine the beams in phase opposition in order to strongly reduce the on-axis starlight while transmitting the flux of off-axis sources located at angular spacings given by odd multiples of 0.5$\lambda/B$ (where $B=14.4$\,m is the distance between the telescope centers and $\lambda$ is the wavelength of observation, see transmission pattern in Figure~\ref{fig:tm}). The high-angular resolution information on the observed object is then encoded in the null depth, which is defined as the ratio of the flux measured in destructive interference and that measured in constructive interference. The advantage of obtaining null depth measurements is that they are more robust against systematic errors than visibility measurements and hence lead to a better accuracy \citep[e.g.,][]{Colavita:2010}.


The LBTI nulling beam combination scheme is represented in Figure~\ref{fig:combination}. The beams are combined in the pupil plane on a 50/50 beamsplitter that can be translated to equalize the pathlengths between the two sides of the interferometer. To achieve an achromatic suppression of light over a sufficiently large bandwidth (8-13\,$\upmu$m), a slight excess of pathlength in one beam is compensated with a suitable thickness of dielectric material in the opposite beam, which allows us to balance the dispersion in the pathlength difference. This technique permits a very simple beam combination while allowing a suitably wide bandpass for starlight suppression. One output of the interferometer is reflected on a short-pass dichroic and focused on the NOMIC camera. NOMIC uses a 1024x1024 Raytheon Aquarius detector that covers a field of view (FOV) of 12$\times$12 square arcsec with a plate-scale of 0.018~arcsec. However, to improve the data acquisition efficiency, only a small portion of the array is read and saved (generally a 256x256 sub-array corresponding to a field-of-view of 4\farcs{6}$\times$4\farcs{6}).

Compared to other ground-based nulling instruments, the LBTI re-images the nulled output of the beamsplitter rather than integrating over a single pixel (or several pixels for dispersed data). Therefore, the LBTI forms an image of the nulled output at the angular resolution of a single aperture with the high-angular resolution information encoded in the flux of the star. This image corresponds to the object brightness multiplied by the transmission pattern of the nuller (see Figure~\ref{fig:tm}, right) and convolved with the point spread function (PSF) of the individual elements (see Figure~\ref{fig:tm}, left). Because the PSF is broader than the transmission pattern and the source is relatively compact, the interference fringes are not visible in the focal plane and images are formed. It is hence possible to study the spatial structure of the detected excesses at the angular resolution of a single aperture. 


\subsection{Differential pathlength and tip/tilt sensing}\label{sec:sensing}

Differential tip-tilt and phase variations between the two AO-corrected apertures are measured with PHASECam, LBTI's near-infrared camera. PHASECam uses a fast-readout PICNIC detector that receives the near-infrared light from both interferometric outputs. The optics provide a field of view of 10$\times$10~arcsec$^2$ with pixels 0.078~arcsec wide and can be adapted to create different setups for pathlength sensing. Three options are currently built into the LBTI to allow a flexible approach to phase sensing: 1) use the relative intensity between the two interferometric outputs, 2) use dispersed fringes via a low-dispersion prism, or 3) use an image of the combined pupils via a reimaging lens. Various neutral density filters are also available together with standard H and K filters. 

\begin{figure}[!t]
	\begin{center}
		\includegraphics[height=4.2 cm]{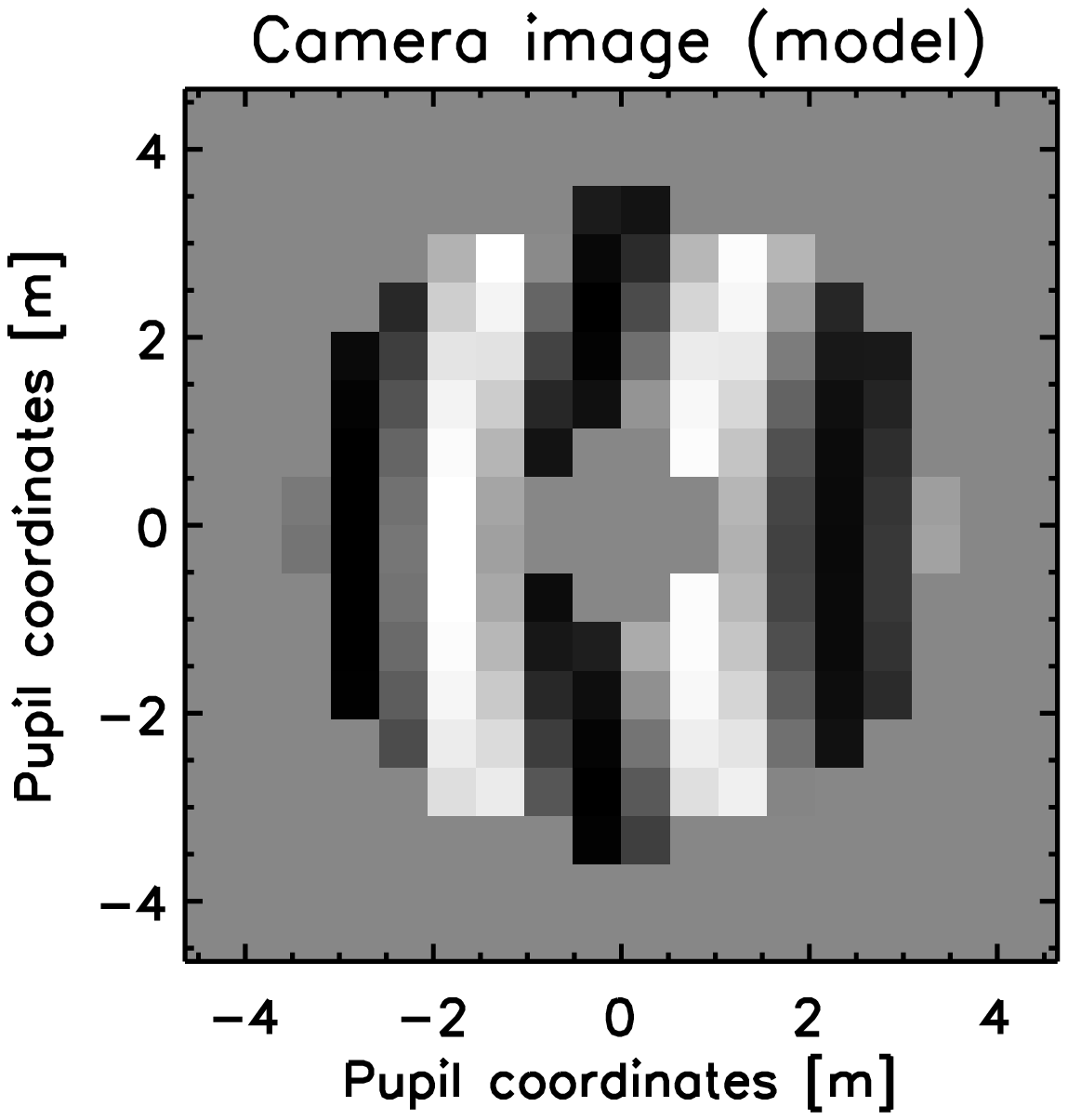}
		\includegraphics[height=4.2 cm]{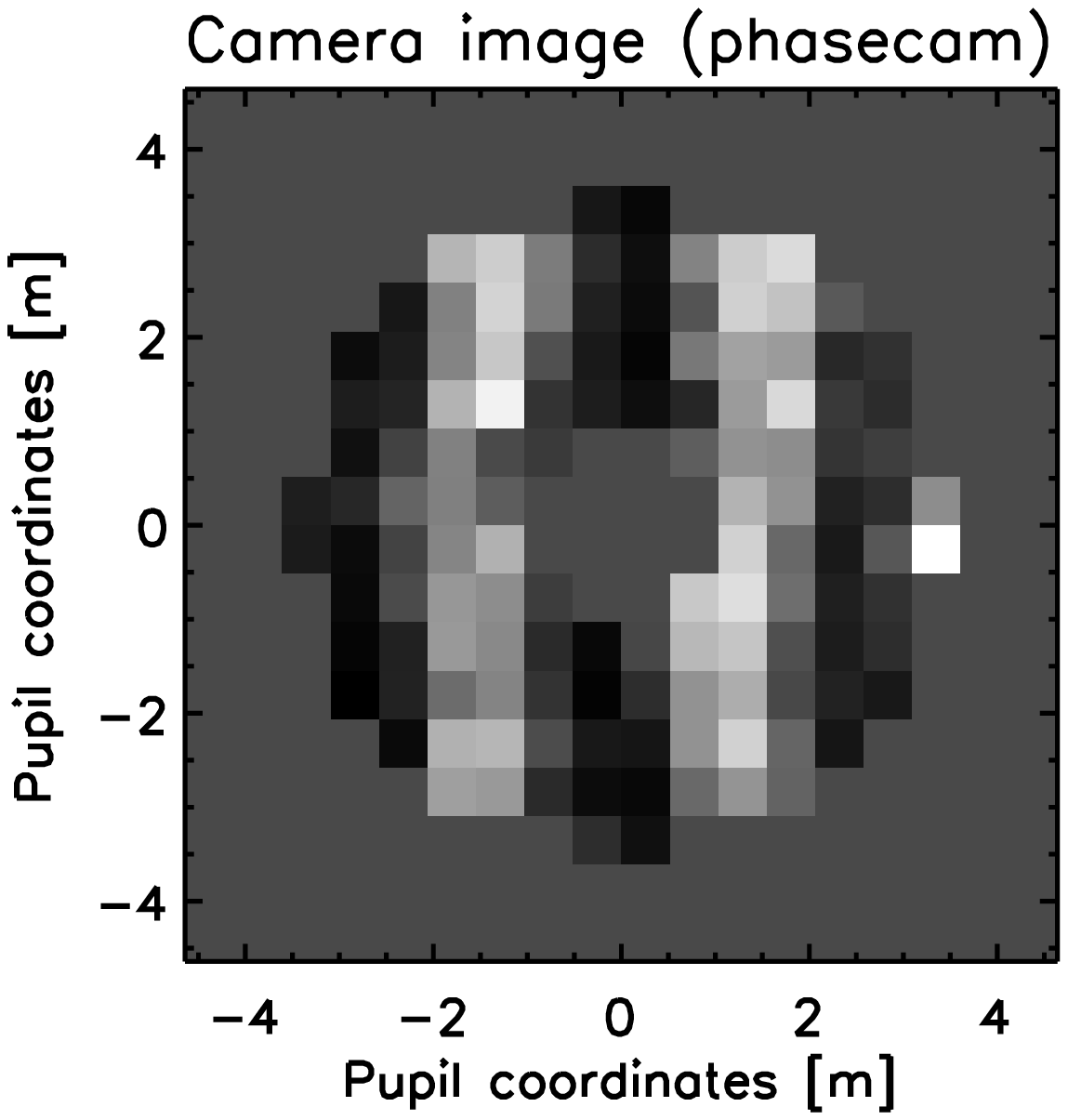}
		\includegraphics[height=4.2 cm]{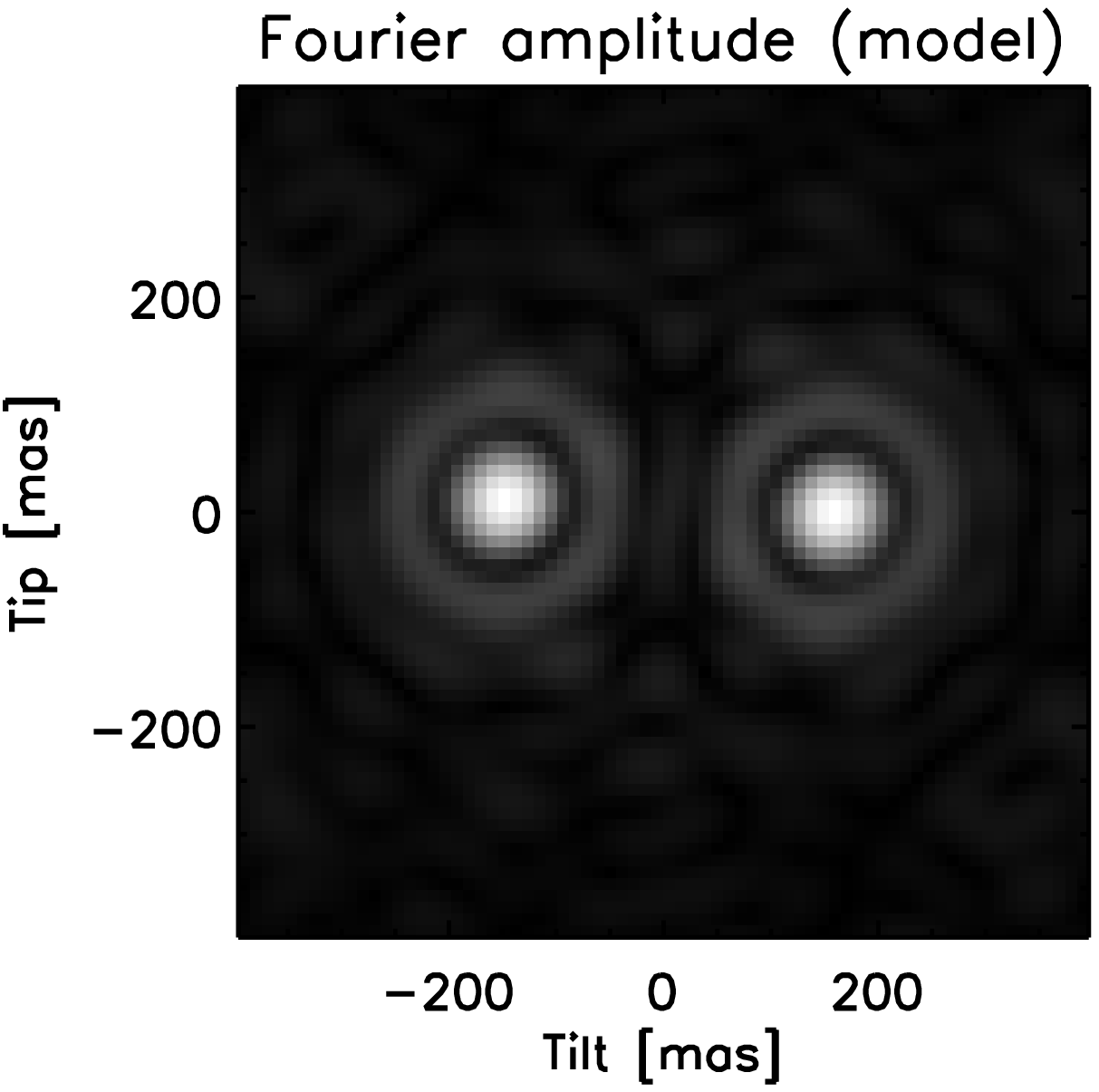}
		\includegraphics[height=4.2 cm]{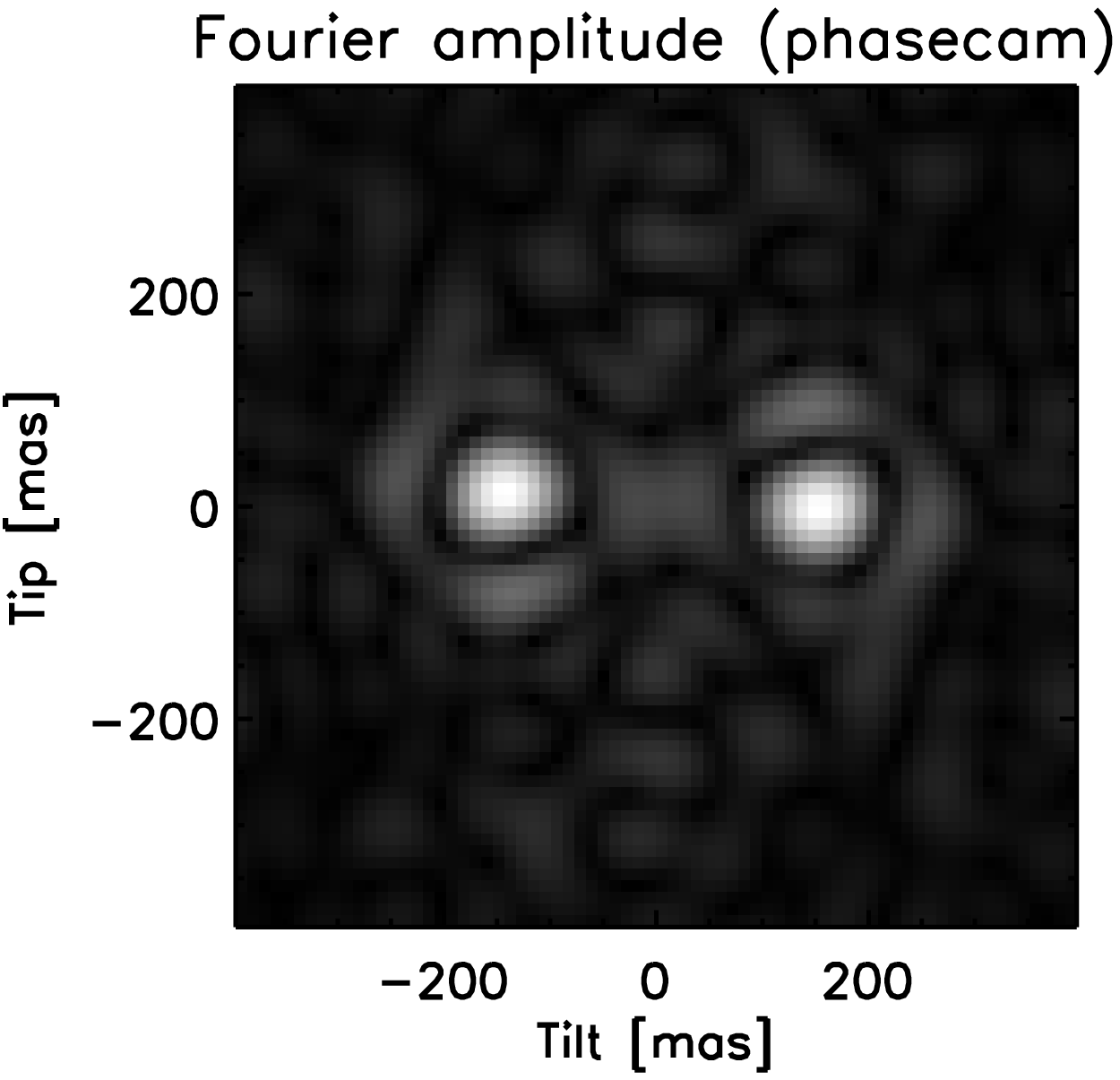}
		\includegraphics[height=4.2 cm]{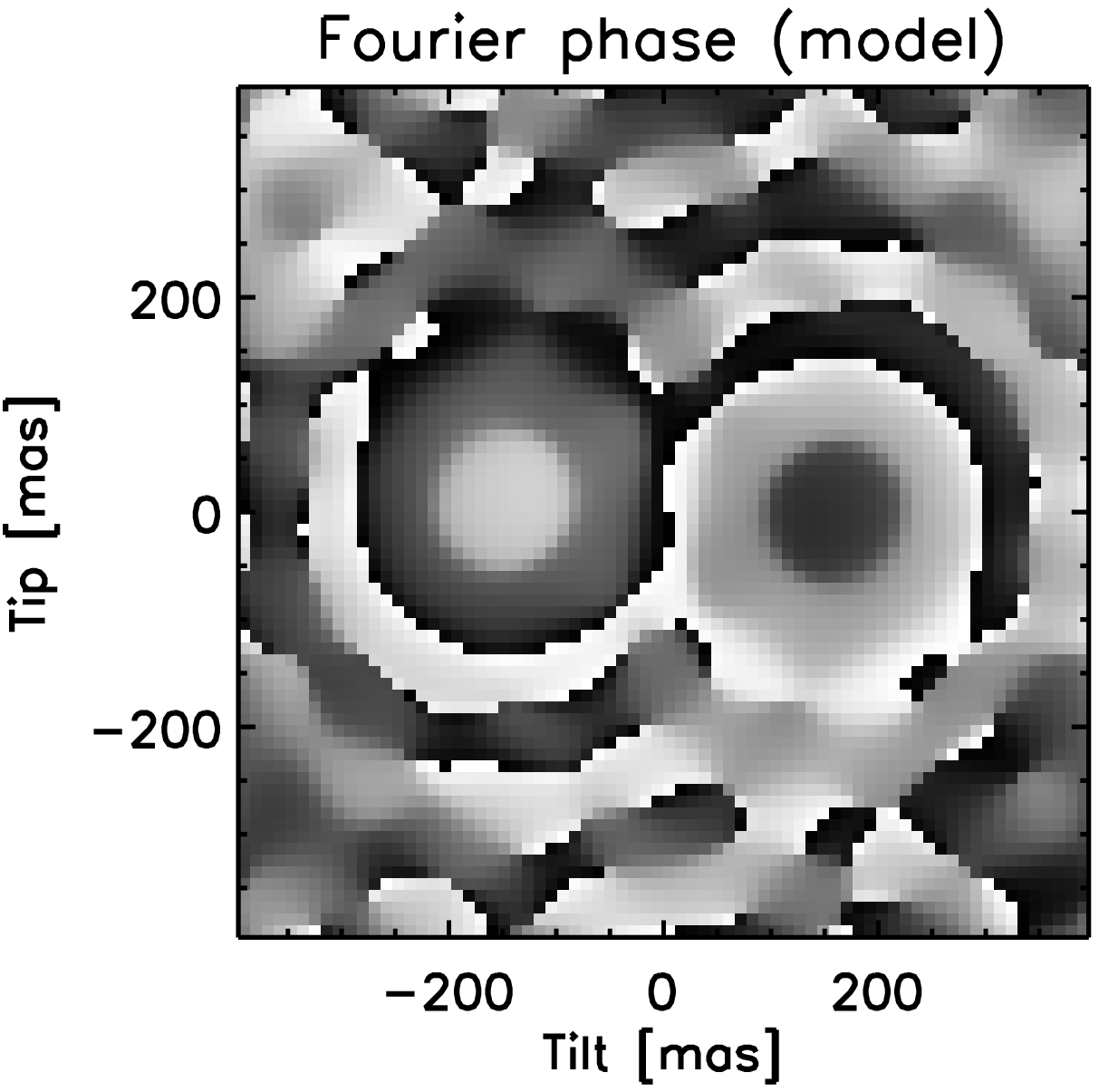}
		\includegraphics[height=4.2 cm]{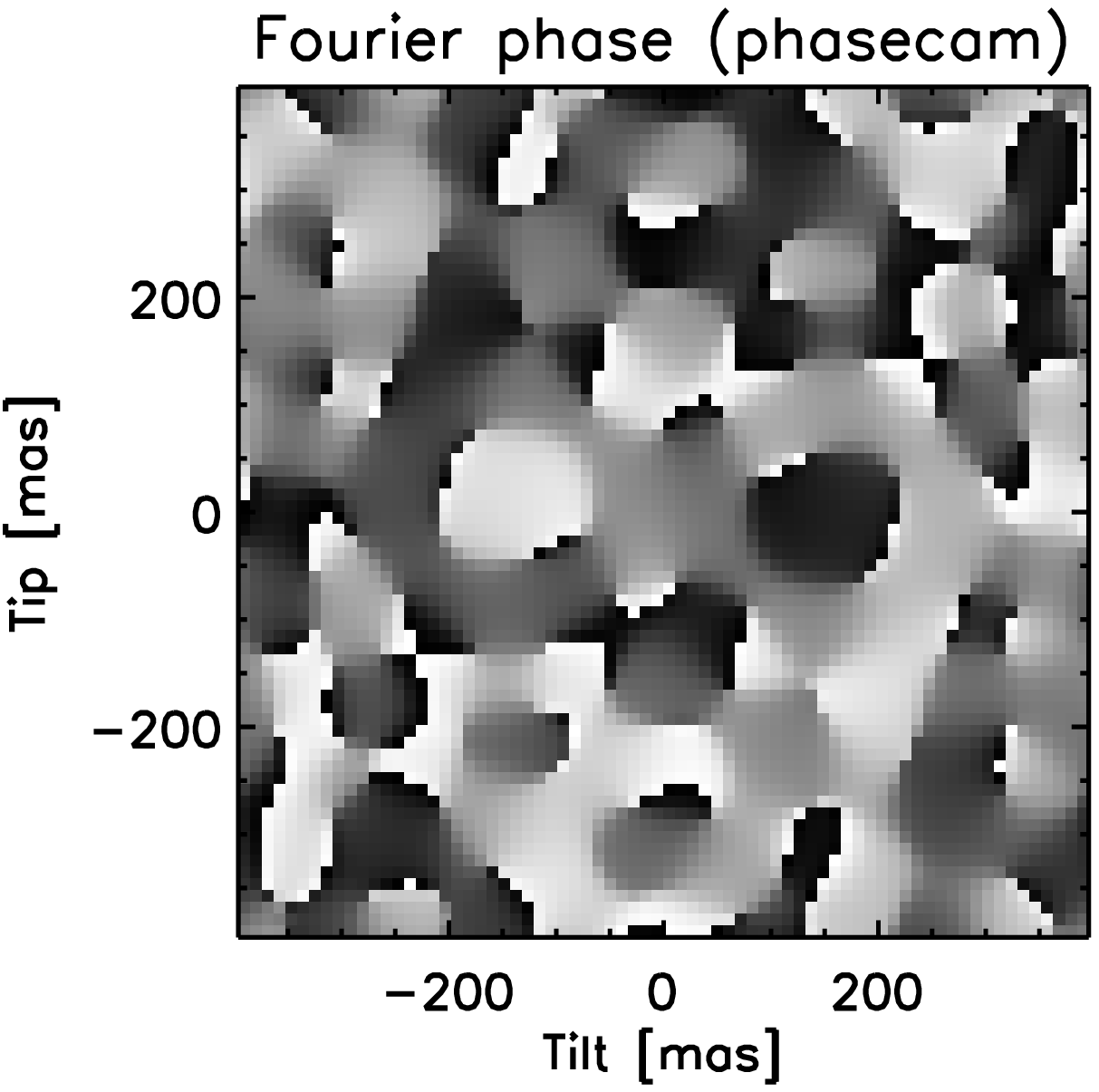}
		\caption{LBTI's phase sensing approach (noise-free model on the left and on-sky K-band data from March 17th 2014 on the right). Pupil images of the two interferometric outputs are formed on PHASECam (one output shown on top) and the Fourier transform is computed to sense both tip/tilt and phase. The peak position in the amplitude of the Fourier image (middle images) provides the differential tip/tilt error signal while the argument of the Fourier image (bottom images) at the peak position provides the phase (grey scale ranging from -$\pi$ to $\pi$). The central region of the pupil is excluded from the FFT (location of the secondary mirrors). Note that the camera image is padded in order to increase the resolution in the Fourier space.} \label{fig:approach}
	\end{center}
\end{figure}

So far, fringe sensing has been mainly performed using a K-band image of pupil fringes (equivalent to wedge fringes). Because of angular dispersion between 2~$\upmu$m and 10~$\upmu$m in the beamsplitter, a well overlapped set of images at 10~$\upmu$m corresponds to a tilt difference of roughly 3 fringes across the pupil at 2~$\upmu$m. This has the nice feature of providing a signal in the Fourier plane well separated from the zero-frequency component and allow us to separate differential tip/tilt and phase variations via a Fourier transform of the detected light.The peak position in the amplitude of the Fourier transform gives a measurement of the differential tip/tilt while the argument of the Fourier transform at the peak position gives a measurement of the optical path delay.  While both outputs are read simultaneously by the detector, only one has been processed for the data presented in this paper. This approach is represented in Figure~\ref{fig:approach} for a noise-free model (left column) and on-sky data from March 17th 2014 (right column). 

\begin{figure}[!t]
	\begin{center}
		\includegraphics[width=8.0 cm]{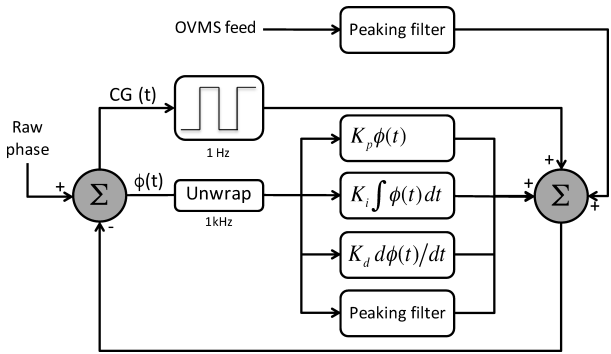}
		\caption{Block diagram of LBTI OPD controller. The measured phase is first unwrapped and then goes through a classical PID controller. A peaking filter is also used to improve the rejection of specific vibration frequencies. An outer loop running at typically 1\,Hz is used to monitor the group delay and capture occasional fringe jumps. In addition, real-time OPD variations induced by the LBT structure are measured by accelerometers all over the telescope (OVMS system) and feedforwarded to the FPC. Differential tip/tilt is controlled following the same principle, except for the unwrapper and the outer loop.}\label{fig:loop}
	\end{center}
\end{figure}

Due to the nature of the Fourier transform measurement, the derived phase is limited to values in the [$-\pi$,$\pi$] range while the phase fluctuations have a typical amplitude of $\sim5$\,$\upmu$m (i.e., $\sim4\pi$ at K band, see Section~\ref{sec:phase}). To get around this issue, the measured phase is unwrapped using a first-order derivative estimate \citep{Colavita:2010c}. The unwrapped phase generally follows closely the phase fluctuations but on-sky verification tests have shown that large phase jumps can occasionally occur and cause fringe jumps (even with the phase loop running at 1\,kHz). To capture eventual fringe jumps, the envelope of interference (or the group delay) is tracked simultaneously via the change in contrast of the fringes. A metric called contrast gradient (CG) has been defined as follows:

\begin{equation}
CG = \frac{\sum_i\mid I_i-\mean{I}\mid(x_i-\mean{x})}{\sum_i I_i};\\
\end{equation}

\noindent where $I_i$ is the intensity of a particular pixel in the pupil and $x_i$ is the coordinate in the horizontal direction of that pixel. The contrast gradient is used to monitor the group delay and detect fringe jumps at a typical frequency of 1-2\,Hz. In the future, we plan to replace this algorithm and use instead the phases measured at two different near-infrared wavelengths (one from each output of the interferometer) to derive the group delay.

Tip/tilt and phase delay tracking are carried out at full speed (i.e., $\sim$1\,kHz) using a classical proportional-integral-derivative (PID) controller (see Figure~\ref{fig:loop}). The PID controller continuously computes the difference between the measured unwrapped phase and differential tip/tilt and their setpoints. Currently, the setpoints are optimized manually by searching for the values that minimize the real-time null depth estimate. The differential tip/tilt setpoint is found to be stable from night to night and is generally checked only once at the beginning of the night. The phase setpoint is checked at the beginning of each observing sequence and adjusted if necessary. The procedure consists in scanning the phase setpoint by increments of 10-20$\degree$ until the minimum real-time null depth estimate is reached. The PID gain parameters are optimized manually to minimize the error signal and generally consist of a large integral gain, a small proportional gain, and no derivative gain. 

\begin{figure*}[!t]
	\begin{center}
		\includegraphics[width=17.5 cm]{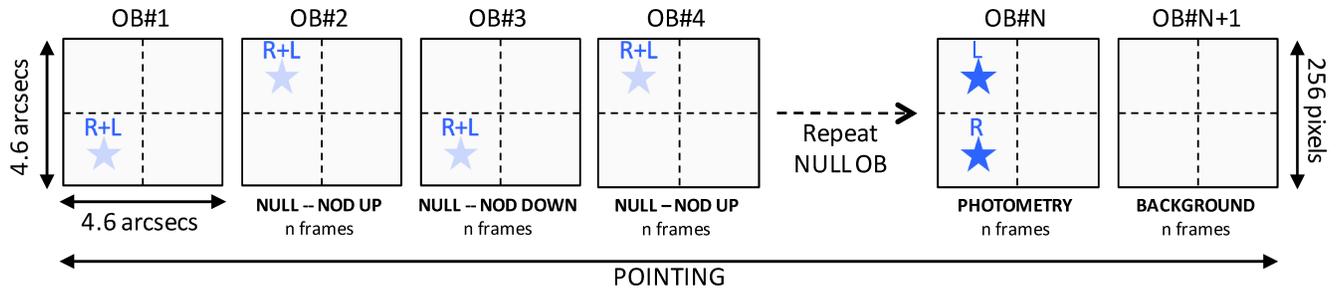}
		\caption{Typical observing sequence used for the HOSTS survey. The sequence is divided in basic observing blocks (OB) consisting of one to two thousand frames, each having an integration time of typically 10 to 100\,ms (depending on the brightness of the star). The complete sequence is composed of several successive OBs at null, i.e.\ with the beams from both apertures coherently overlapped in phase opposition, one OB of photometric measurements with the beams separated on the detector, and one OB of background measurements with the beams nodded off the detector. Each square represents a 256 x 256 subframe of the NOMIC detector that covers a region of approximately 4\farcs{6} x 4\farcs{6}. The beams are aligned vertically in the middle of a given channel (see blue stars) to maximize the effective field of view and nodded back and forth by 2\farcs{3} (up-down to preserve the differential pathlength). The dashed lines represent the limits of different detector channels. The right channels of the detector are not used directly for the null depth measurements but are useful for diagnostics and frame selection.}\label{fig:sequence}
	\end{center}
\end{figure*}

In addition to the real-time control described above, OPD and tip/tilt vibrations induced by the telescope structure, which are measured in real-time by accelerometers all over the telescope \citep[OVMS system,][]{Kurster:2010}, are feedforwarded to the FPC. For the data presented in this study, we only used the accelerometers located on the secondary mirrors, which produce significant OPD variations at a frequency of $\sim$12\,Hz \citep[typically a few hundred nanometers RMS,][]{Defrere:2014}. Peaking filters are also used to improve the rejection of specific vibration frequencies not captured by the feedfoward system (e.g., at $\sim$15Hz due to the tertiary mirrors). They are biquad filters applied directly to the control algorithm that drives the FPC. Finally, note that PHASECam does not capture phase and differential tip/tilt variations that are variable between the near-infrared, where it operates, and the mid-infrared, where the null depth measurements are obtained (see more information in Section~\ref{sec:phase}). Commissioning observations have shown that the loop can run at full speed down to a magnitude of K$\sim$6.5, which is sufficient to observe all targets of the HOSTS survey sample. The loop frequency can in principle be decreased down to $\sim$30Hz to observe fainter objects but this has never been tested and requires more investigation. 


\section{Data reduction}\label{sec:reduction}
\subsection{Level 0 Data Products}

The raw data from the nuller (Level 0) consist of single detector frames saved in the {\sc fits} format. Each fits file consists of an image, which contains generally 256 x 256 pixels, and header information, which contains approximately 160 keywords and associated data. These keywords are grouped according to their origin and include for instance the target information (e.g., name, RA, DEC), the telescope telemetry (e.g., elevation, azimuth), the AO telemetry (e.g., loop status, loop frequency, loop gains), the detector and filter information (e.g., integration time, filter position, gain), the PHASECam telemetry (e.g., loop status, loop frequency, measured phase and differential tip/tilt), and weather information (e.g., seeing, wind). The science frames are acquired according to a pre-defined observing sequence as shown in Figure~\ref{fig:sequence}. The basic observing block (OB) consists of one to two thousand frames, each having an integration time of typically 10 to 100\,ms, depending on the brightness of the star. The observing sequence is composed of several successive OBs at null, i.e.\ with the beams from both apertures coherently overlapped in phase opposition, one OB of photometric measurements with the beams separated on the detector, and one OB of background measurements with the beams nodded off the detector. In order to estimate and subtract the mid-IR background, the OBs at null are acquired in two telescope nod positions separated by 2\farcs{3} on the detector (see background subtraction strategy in Section~\ref{sec:background}). All successive OBs on the same target define a pointing and the plan for the HOSTS survey is to acquire three pointings on the science object, interleaved with pointings on reference stars to measure and calibrate the instrumental null floor (e.g., CAL1-SCI-CAL2-SCI-CAL1-SCI-CAL3 sequence). To minimize systematic errors, calibrator targets are chosen close to the science target, both in terms of sky position and magnitude, using the \emph{SearchCal} software \cite[][see Appendix~\ref{app:calib} for more information about calibrator selection]{Bonneau:2011}. 

\subsection{L0 image reduction}\label{sec:l0red}

Reduction of detector frames consists of several well-defined steps that generally include bias subtraction, bad pixel removal, and flat fielding. For our mid-infrared observations, these steps are either not necessary or achieved by nodding subtraction. For instance, the detector bias is removed via nodding subtraction occurring every few minutes at most, which is fast enough not to worry about variability due to temperature variations (related to the telescope elevation). Flat fielding can in principle be derived from observations of the sky at different airmasses (using for instance the background OB of each pointing, see Figure~\ref{fig:sequence}). However, it is generally not possible to find a satisfactory method of creating image flats that do not also significantly increase the noise level in the images. Besides, the detector response is intrinsically flat over the whole detector, of which we are only interested in a very small region ($\sim$ 50 x 50 pixels for a typical HOSTS star). In addition, thanks to some flexibility in the alignment of the instrument, this region is chosen to be very clean and contains very few bad pixels. While the reduction software has the capability to perform these steps, they are generally not executed and the L0 image reduction only consists in grouping the frames of the same nod position in a single data cube for faster data access in the following step. 

\subsection{Background subtraction}\label{sec:background}

A critical step for obtaining accurate null depth measurements is to correctly estimate and subtract the background level at the position of the nulled image. In the mid-infrared, this is a challenging task due to the strong, non-uniform, and rapidly varying background emission (see top panel of Figure~\ref{fig:multi_bckg}). Under typical observing conditions and operating parameters for the LBTI in nulling mode, the total instrumental emissivity in the N' band (10.22-12.49\,$\upmu$m, see transmission curve in Appendix~\ref{app:trans}) is approximately 27\% when the beam combiner is warm, which was the case for all the observations presented in this paper. Including the atmosphere, the total thermal background is equivalent to a 240\,Jy source over a photometric aperture with a diameter of 0\farcs{28} (LBT PSF diameter). This is 3200 Jy/arcsec$^2$, or 1 Jy/pixel. This is nearly a million times brighter than the faintest detectable level of zodiacal dust planned for the LBTI survey. To reach background-limited noise performance, the background is estimated in a two step process: the first step uses simultaneous background measurements obtained in an annulus around the position of the beams while the second step estimates the background at the same position as the beams but at a different time (after the telescope has been nodded). The first step deals with background time variations, which mostly originate from the atmosphere, while the second step deals with the non-uniformity of the background across the detector, which is mostly due to the instrument. These steps are described in more detail as part of the flux computation in the next section.

\subsubsection{Flux computation}\label{sec:flux}

Flux computation is performed in every frame by circular aperture photometry using the {\tt aper.pro} IDL astrolib routine. We use an aperture radius of 0.514$\lambda$/D, where D is the diameter of the primary aperture. This is equivalent to a radius of 140\,mas (or 8 pixels) at 11.1\,$\upmu$m and an area of 201 pixels. Because the flux of the star at null is usually too faint for precise beam centroid determination in a single frame, the beam centroid is computed on the median-combined image of all consecutive frames of the same nod position. We apply a double-pass centroid function that applies first a coarse Gaussian fit to the whole detector frame to find the approximate beam position and then uses the non-linear least squares fitter MPFIT \citep{Markwardt:2009} around this position for sub-pixel accuracy. Aperture noise resulting from the intersection of the circular aperture with square pixels is taken into account in {\tt aper.pro}, which computes the exact fraction of each pixel that falls within the photometric aperture.

\begin{figure}[!t]
	\begin{center}
		\includegraphics[width=8.5 cm]{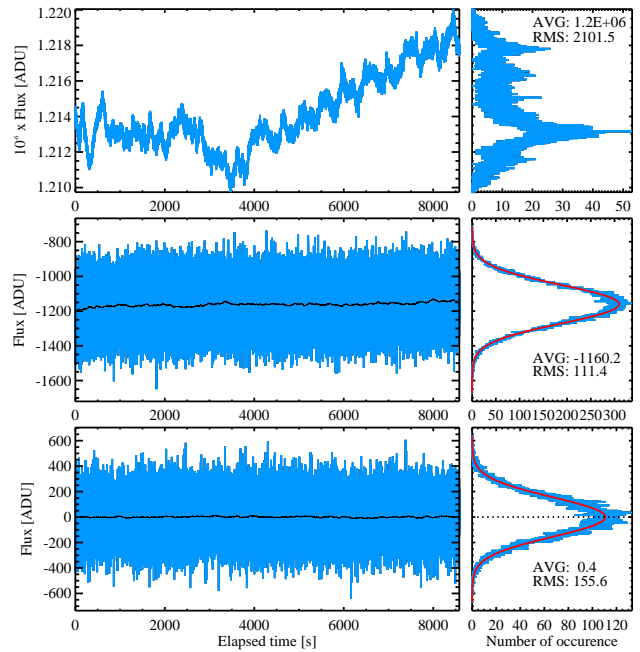}
		\caption{Top: example of on-sky raw thermal background measurements obtained in the N' band with the telescope pointing at an empty region of the sky and covering approximately 15\,degrees of elevation change during the whole duration of the sequence. The left panel shows the flux integrated over a photometric aperture of 8 pixels in radius while the right panel shows the corresponding distribution. Middle: same measurements after subtraction of simultaneous background measurements (left). The corresponding distribution (right) is now Gaussian and shows a relatively large offset. The black line represents a running average of 100 seconds to better show the low-frequency drift due to slowly-changing instrumental background. Bottom, same measurements after subtraction of simultaneous background measurements and nod subtraction (left). For this example, nod subtraction has been performed at the maximum frequency (i.e., using adjacent frames). The corresponding distribution (right) is now Gaussian and centered on 0. These data have been obtained using an integration time of 28\,ms (on May 14, 2014).}\label{fig:multi_bckg}
	\end{center}
\end{figure}

Aperture photometry has the advantage to remove the real-time background fluctuations using a circular annulus centered around the photometric aperture. The sigma-clipped average of all pixels in the background annulus is computed and subtracted on a per-pixel basis from each pixel in the photometric aperture. To minimize photon and readout noise, the inner radius of the background regions is generally chosen as close as possible to the photometric aperture but can be extended for bright nearby stars for which the habitable zone is extended and photometric errors are not dominant. The outer radius of the regions is generally chosen as the smallest distance to a channel edge but can be adapted in presence of an obvious background bias (see Appendix~\ref{app:bckg} for more information). The result of this process is shown in Figure~\ref{fig:multi_bckg} (see middle panel), where most of the background fluctuations are now corrected. There remains however a slow drift around a relatively large negative value. This offset between the background region and the photometric aperture comes from the spatial structure of the background while the drift is due to slowly-moving optics inside the LBTI (related to the telescope elevation). These effects must be corrected or they will create a flux-dependent background bias between stars of different magnitudes.

To correct for this offset and slow drift, we apply a second step of background subtraction using a median-combined image of frames in adjacent nods. Aperture photometry is applied at the exact same position on the detector (now pointing to an empty region of the sky) and the resulting flux is subtracted from the flux of the star at null. Because of the slow drift, it is important to minimize the nodding period so that we only use the $n$ closest frames in time of the adjacent nods, where $n$ is the number of frames in the current nod. The result obtained with both aperture photometry and nod subtraction is shown for the maximum nodding frequency in the bottom panel of Figure~\ref{fig:multi_bckg}, where the residual offset is now close to 0 and no low-frequency drift is visible. In practice, there is a trade-off between observing efficiency and nodding frequency. If everything goes smoothly, it takes generally 5 to 10 seconds to close the AO and phase loops so that we stay at least 60 to 90 seconds in a given nod position. This is sufficiently long for the background to change between the two nod positions due to slowly-moving optics inside the LBTI. Therefore, the background illumination is not necessarily perfectly uniform after nodding subtraction, which creates a small offset between the background estimated from the background annulus and that in the photometric region. This offset, called background bias in the following, is present in most of our commissioning data and can be up to ten times larger than the photometric noise. New alignment procedures will be developed during the science validation phase to mitigate this bias.

  
\subsection{Null depth computation}\label{sec:null}

The last step in the L0 data reduction is to convert the flux measurements at null of each OB to single values and corresponding error bars. The classical way to do that is to compute the average or mode of the null depth measurements, i.e.\ the flux measurements at null divided by the constructive flux $I_+$:
\begin{equation}
	I_+ = I_1 + I_2 + 2\sqrt{I_1I_2}\, , \label{eq:cons}
\end{equation}
where $I_1$ and $I_2$ are the mean individual intensities measured during the same pointing. While straightforward to implement, the classical technique is also very sensitive to instrumental imperfections that vary between the calibrator and the science stars. In the case of the LBTI, a major error term comes from the mean phase setpoint, which varies from one OB to the next and is indistinguishable from a true extended emission with classical reduction techniques. This setpoint offset appears for two reasons. First, the fringe tracker computes the Fourier transform of the combined pupil image by defining a circle around the illuminated portion of the detector. This circle is only defined to a precision of one pixel at the beginning of each OB at null so that any sub-pixel drift of the beams is interpreted as a setpoint change. Since there are approximately 4 pixels per fringe, one pixel corresponds to a pathlength offset of $\sim$0.5\,$\upmu$m (or $\sim$0.3\,rad at 11\,$\upmu$m). Second, the water vapor component of the atmospheric seeing creates a variable differential phase between the K band, where the phase is measured and tracked, and the N band, where the null depth measurements are obtained. Whereas the phase setpoint is optimized regularly to minimize this effect, the water vapor component of the atmospheric seeing can be sufficiently fast (typically of the order of a few seconds) to modify the mean phase setpoint during the acquisition of a single OB. 

To get around this issue of varying mean phase setpoint, we use the statistical reduction (or NSC) technique mentioned in the introduction. This technique can derive the mean phase setpoint from the null depth measurements themselves and has demonstrated improved calibrated null accuracies over those obtained with classical reduction techniques in the case of the PFN (up to a factor of 10). The idea behind NSC is to produce a synthetic sequence of flux measurements at null and compare its distribution to that of the measured sequence. The synthetic instantaneous flux sequence $I\_(t)$ is created using the following expression \citep[e.g.,][]{Serabyn:2000,Mennesson:2011}:
\begin{equation}
	I\_(t) = I_1(t) + I_2(t) + 2 |V|\sqrt{I_1(t)I_2(t)}\cos(\Delta\phi(t)) + B(t)\, , \label{eq:null}
\end{equation}
where $I_1$(t) and $I_2$(t) are the individual instantaneous photometries, \(\lvert V\rvert\) is the absolute value of the source visibility at the instrument baseline (which is related to the source null as explained later in this section), $\Delta\phi$(t) is the instantaneous phase offset (close to $\pi$ at null), and $B(t)$ is the instantaneous measured background. The advantage of using distributions is that it is not necessary to have access to simultaneous auxiliary measurements to create the synthetic flux distributions. The distribution of the unknown instantaneous sequences (i.e., $I_1$(t), $I_2$(t), and $B(t)$) are estimated by their values measured at a slightly different time. The observing sequence presented in Figure~\ref{fig:sequence} has been specifically defined for that purpose. Given their excellent stability, the photometric intensities ($I_1$ and $I_2$) are obtained only once per pointing, generally at the end of the sequence. The background measurements $B(t)$ on the other hand are estimated at a higher frequency and usually in the closest adjacent nod (and at the same position as the null depth measurements). Note that both measured photometric intensities are intrinsically affected by the photon noise of the thermal background so that their distributions do not represent the true intensity variations. Because the background noise is overwhelmingly dominant and already included in the term $B(t)$, there are no reasons to inject the measured distributions of $I_1$ and $I_2$ in the model. Instead, we use two constants derived by averaging the photometric measurements of each aperture.

\begin{figure}[!t]
	\begin{center}
		\includegraphics[width=8.0 cm]{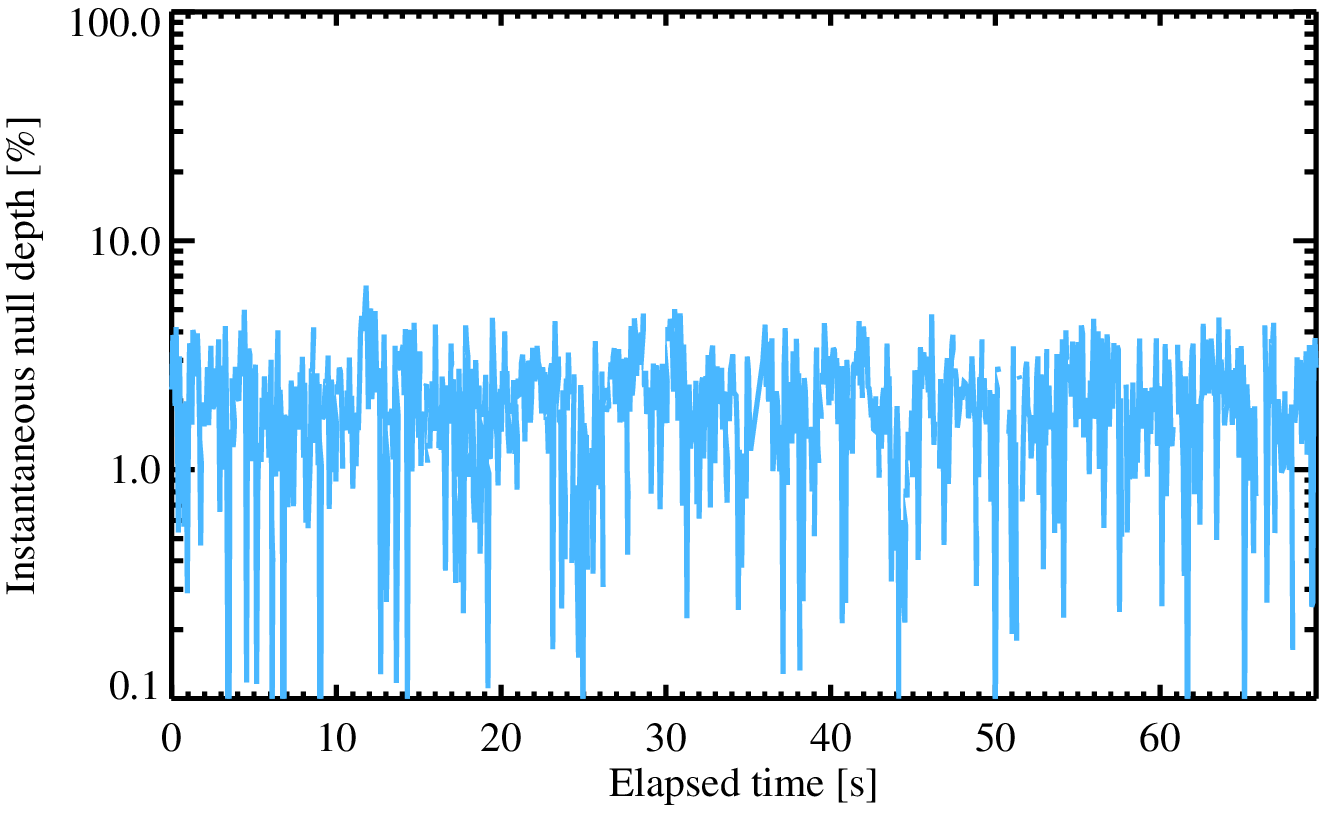}
		\includegraphics[width=8.0 cm]{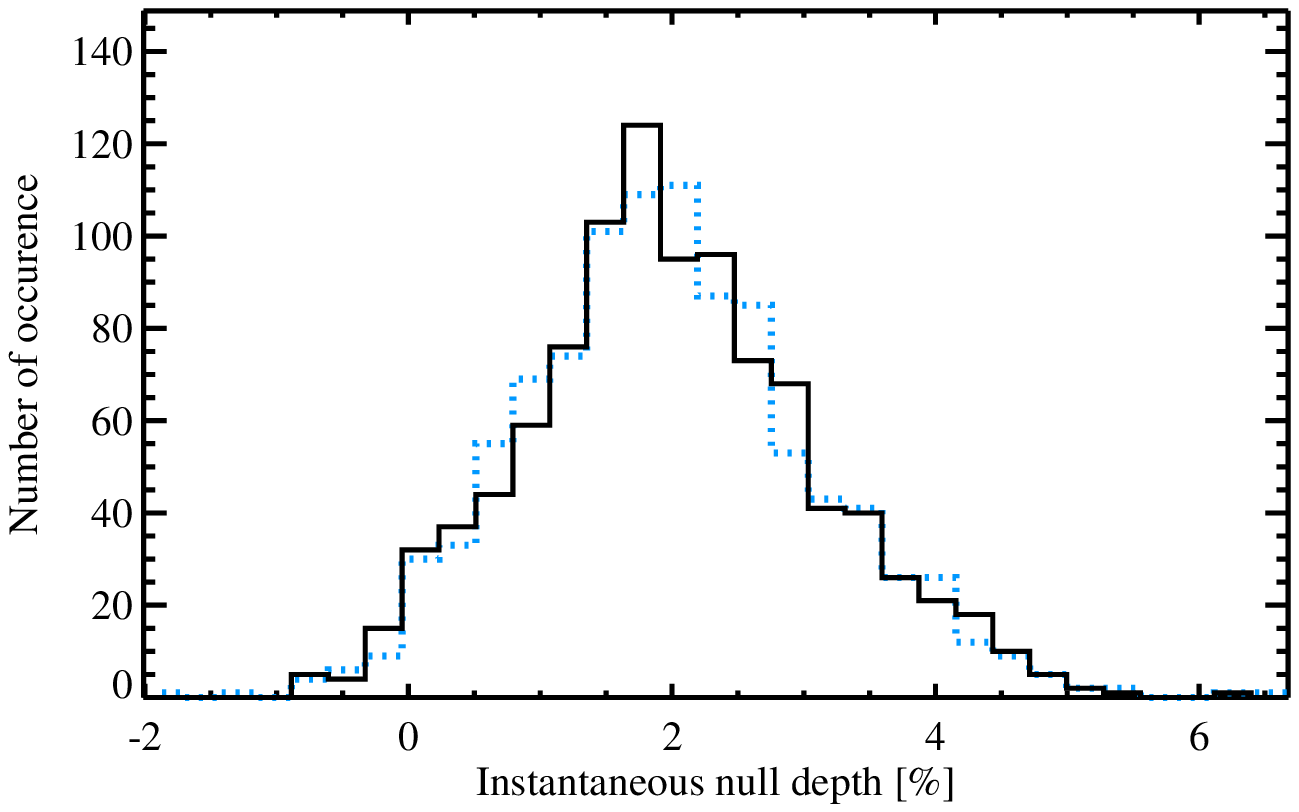}	
		\caption{Top: null depth sequence obtained on $\beta$ Leo (typical results, 2015 February) vs elapsed time in seconds. The large excursions in the null depth are dominated by variable precipitable water vapor. Bottom: measured null depth distribution (solid line) and best-fit synthetic null depth distribution (dashed line). The reduced $\chi^2$ ($\chi^2_r$) amounts to 0.56 (see definition in \citealt{Hanot:2011}, Equation 17).}\label{fig:nullseq}
	\end{center}
\end{figure}

As discussed by \cite{Hanot:2011}, Equation~\ref{eq:null} is only valid for instantaneous flux measurements or if the time-dependent quantities do not vary within each integration time. In the case of the LBTI, the instantaneous differential phase varies significantly at high frequency (see Section~\ref{sec:perfo}), which would require integration times prohibitively short to ``freeze'' it. While integration times as short as 3\,ms could be used in practice, this would lead to a significant sensitivity loss due to readout noise and camera overheads. Therefore, we have modified the synthetic null depth expression to include the effect of varying differential phase over a finite integration time. The average of the flux at null over an integration time $T$ can be expressed from Equation~\ref{eq:null} as:
\begin{equation}
\begin{aligned}
	\mean{I\_(t)} = & I_1 + I_2 + 2 |V|\sqrt{I_1I_2}\mean{\cos(\Delta\phi(t))} + B \\
	          		    = & I_1 + I_2 + 2 |V|\sqrt{I_1I_2}[\cos(\Delta\phi_0)\mean{\cos \epsilon(t)} \\
	                   & -\sin(\Delta\phi_0)\mean{\sin \epsilon(t)}] + B \\
	           \simeq & I_1 + I_2 + 2 |V|\sqrt{I_1I_2}\cos(\Delta\phi_0)(1-0.5\sigma_\epsilon^2) + B\, , \label{eq:null3}
\end{aligned}
\end{equation}
where $I_1$, $I_1$, and $B$ are respectively the average of $I_1(t)$, $I_2(t)$, and $B(t)$ over the same integration time $T$. The term $\cos(\Delta\phi(t))$ has been replaced by $\cos(\Delta\phi_0+\epsilon(t))$, where $\Delta\phi_0$ is the mean differential phase over $T$ and $\epsilon(t)$ is the instantaneous phase (within $T$). The standard deviation over time $T$ of the phase, $\sigma_\epsilon$, is measured in real-time by PHASECam and recorded in the NOMIC {\tt fits} header. At high-frequency (typically $>$1/60ms $\simeq$ 16\,Hz), the contribution of variable water-vapor to the differential phase is negligible compared to other sources of vibrations and the near-infrared measurement is a very good approximation of the differential phase in the mid infrared. The phase conversion between the two wavebands is achieved by computing the effective wavelength using the spectrum of the star \citep[using tabulated K-band spectra from][]{Pickles:1998} and the spectral transmission of PHASECam. The resulting phase is then used to produce synthetic flux sequences following Equations~\ref{eq:null} and \ref{eq:null3}. Note that in the future, we plan on using the terms $\mean{\cos \epsilon(t)}$ and $\mean{\sin \epsilon(t)}$, which were not yet available at the time of the observations presented in this paper. 

\begin{figure}[!t]
	\begin{center}
		\includegraphics[width=7.3 cm]{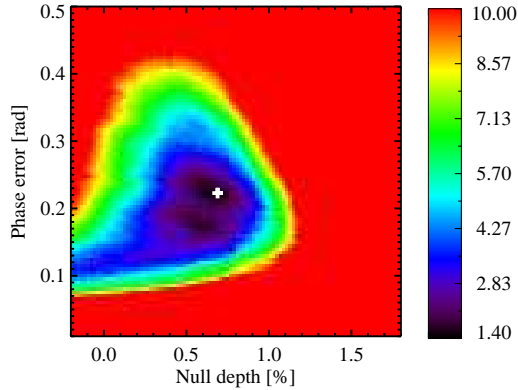}
		\caption{Example of $\chi^2_r$ map represented as a function of null depth and RMS phase error $\sigma_\phi$ ($\chi^2_r$ minimized along the mean phase direction). The region where $\chi^2_r>10$ has been set to 10 to emphasize the low $\chi^2_r$ region. The position of the best-fit model is represented by the white cross.}\label{fig:nullmap}
	\end{center}
\end{figure}

Replacing the visibility by the null depth as \citep{Mennesson:2011}:
\begin{equation}
	N = \frac{1-|V|}{1+|V|}\, , \label{eq:vtonas}
\end{equation}
and using the auxiliary data ($I_1(t)$, $I_2(t)$, $B(t)$, and $\sigma_\epsilon^2(t)$), a synthetic flux sequence can be produced by fixing the three remaining parameters in Equation~\ref{eq:null}, i.e. $N$, $\mu_\phi$ (the mean value of $\Delta\phi(t)$ over all the measurements of the OB), and $\sigma_\phi$ its standard deviation. The method then consists in creating a large grid of models over these three parameters and comparing them to the measured sequence. The best-fit parameters are derived using a least square estimator (i.e., by $\chi^2$ minimization) applied to the whole grid. In theory, the least square estimator is equivalent to the maximum likelihood estimator only in the presence of Gaussian noise. This is a valid assumption in our case because the noise terms are $n$ independent binomial variables (where $n$ is the number of bins in the histogram), which follow closely a Gaussian distribution if the number of occurrences per bin is sufficiently large. For this reason, we only keep the bins that have at least 10 occurrences in the histogram fit. Note that it is possible to derive a maximum likelihood estimator that does not rely on this assumption. The derivation of this estimator is beyond the scope of this analysis and will be presented in a forthcoming paper. 

Regarding the error bars on the best-fit parameters, they are derived by bootstrapping \citep[independent of the actual noise properties,][]{Efron:1979}. This means that, for each position of the grid, we create a large number (2000) of ``alternative'' flux sequences by randomly resampling and replacing the observed flux measurements at null. Each alternative sequence is compared to the synthetic null depth histogram and the distribution of best-fit null depths gives the statistical uncertainty (68.3\% interval). To avoid running the fit on prohibitively large grids and given the required precision on the  source null, we first run a coarse search to find good starting values for the three parameters and then perform a fine search around these values. 

\begin{figure*}[!t]
	\begin{center}
		\includegraphics[width=8.2 cm]{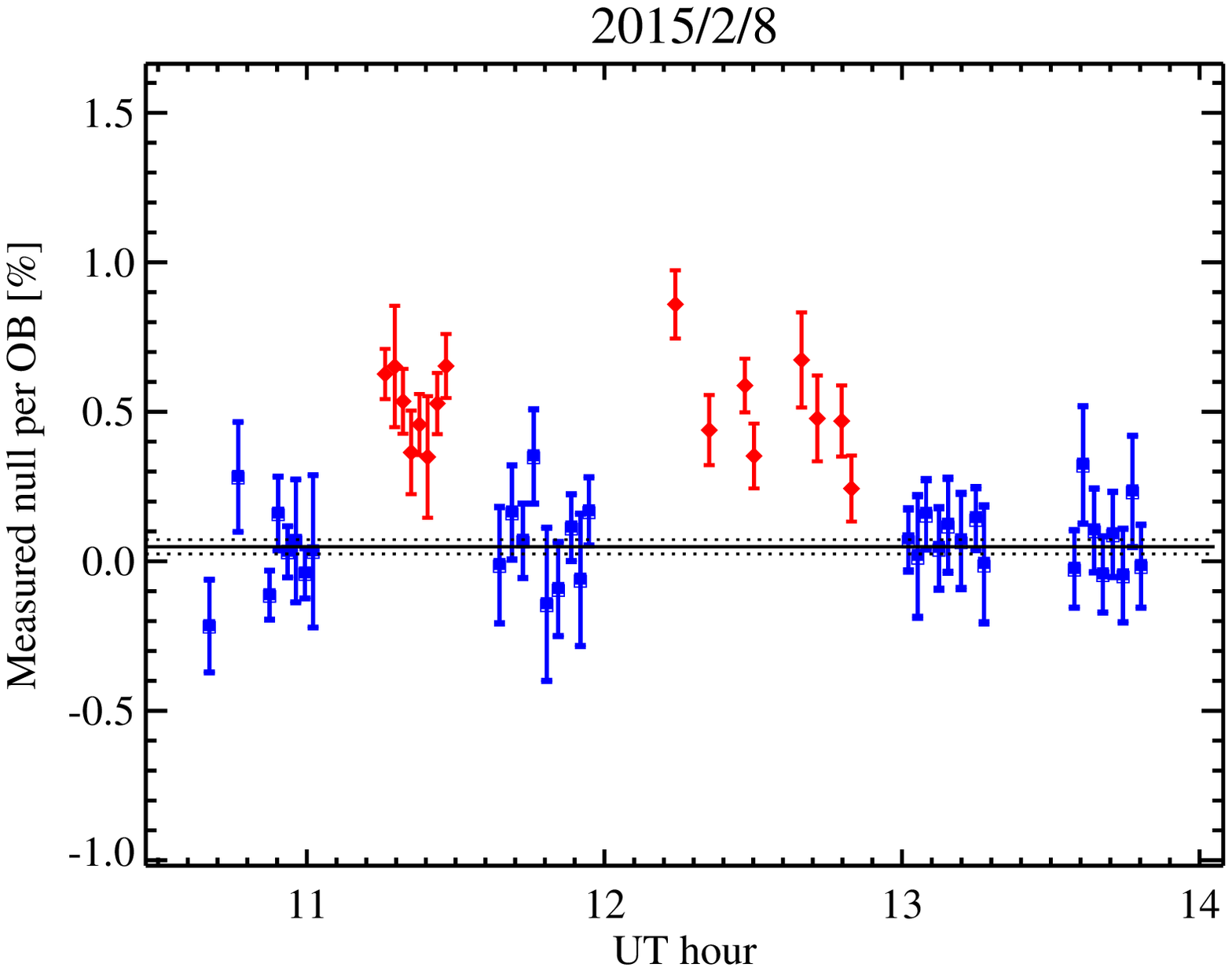}
		\includegraphics[width=8.2 cm]{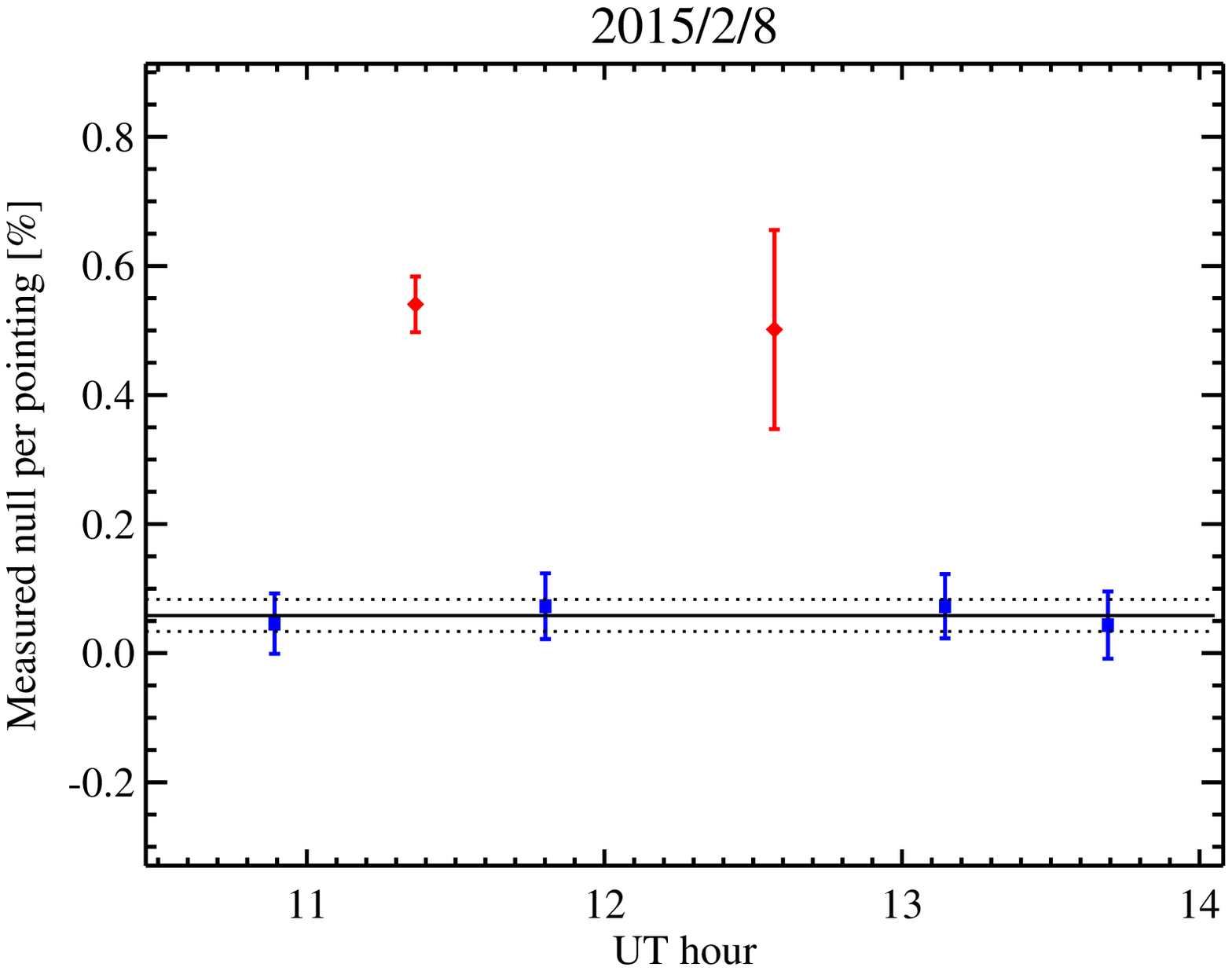}
		\caption{Left, null depth measurements per OB as a function of UT time obtained on February 8, 2015. The blue squares show the calibrator measurements while the red diamonds represent the $\beta$\,Leo measurements. The estimated instrumental null floor is represented by the solid black line and the corresponding 1-$\sigma$ uncertainty by the dotted lines. Right, corresponding null depth measurements per pointing (same notation). The longer time spent to acquire the second $\beta$\,Leo pointing increased the background bias and hence the dispersion of the null depth measurements per OB in the left-hand plot. This effect results in a larger systematic error for this pointing (see Table~\ref{tab:nullp}) and explains the larger error bar in the right-hand plot.}\label{fig:null}
	\end{center}
\end{figure*}

\begin{figure}[!t]
	\begin{center}
		\includegraphics[width=8.2 cm]{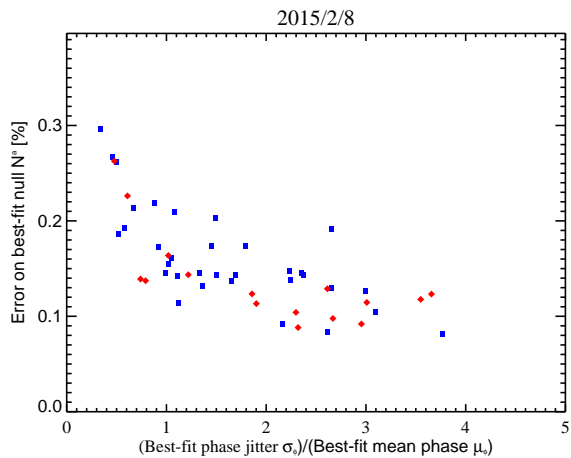}
		\caption{Error on the best-fit null depth versus the ratio between the best-fit phase jitter ($\sigma_\phi$) and the best-fit mean phase ($\mu_\phi$) (same data as in Figure~\ref{fig:null}). The error bar increases with lower values of this ratio as expected from performing the NSC reduction on a finite number of measurements \citep[see Figure~5 in][]{Hanot:2011}.}\label{fig:sigmu}
	\end{center}
\end{figure}

An example of null depth measurements obtained in the N' band is shown in the top panel of Figure~\ref{fig:nullseq}. The corresponding measured and best-fit synthetic null  depth distributions are shown in the bottom panel. Figure~\ref{fig:nullmap} represents the corresponding parameter grid projected on the mean phase direction. The null depth measurements and error bars for each OB of the $\beta$~Leo sequence obtained on February 8, 2015, are shown in Figure~\ref{fig:null} (left). Interestingly, the error bar on the best-fit null depth decreases with the ratio between the best-fit phase jitter and the best-fit mean phase as shown in Figure~\ref{fig:sigmu}. This effect, which may seem counterintuitive, is actually expected from performing the NSC reduction on a finite number of measurements \citep[see Figure~5 in][]{Hanot:2011}. It can be understood by realizing that the NSC is unable to distinguish between a mean phase offset and a true extended emission in the absence of phase jitter. The parameters constraining the minimum number of measurements per OB required to accurately retrieve the best-fit null depth are known (mainly the background noise, the phase jitter, and the mean phase offset) but further investigations are needed to define their sweet spots for the LBTI. 


\subsection{Data gating}

All the processing steps described in the previous sections use data that have passed through various data-quality checks, which vary depending on the nature of the frame (null, photometry, or background frame). The first obvious gating consists of the removal of all open-loop frames using the AO and PHASECam telemetry. For photometric and null frames, we require both AO systems to be in closed-loop. For null frames, we require in addition that the fringe sensor is in closed loop, that there were no fringe jumps during the previous integration time of NOMIC, and that the near-infrared fringe quality (tracked by the SNR of the fringes) is not poorer than a certain level (3-$\sigma$ threshold). Since fringe jumps are only monitored at 1\,Hz (see Section~\ref{sec:sensing}), we also gate the null frames by removing those showing a null depth larger than 20\%. The null depth is computed directly by dividing the flux measurements at null by the constructive flux estimated using Equation~\ref{eq:cons}. Finally, the last data gating applies to all kind of frames and consists in removing the frames which show a high background level (5-$\sigma$ threshold). These criteria generally remove less than $\sim1\%$ of the frames under typical conditions with no major loop failure. Figure~\ref{fig:nullseq} shows an example of gated null depth measurements (top panel) and corresponding distribution (bottom panel). The null depth typically fluctuates around a few percents with low-frequency variations due to precipitable water vapor (PWV, see Section~\ref{sec:phase}).

\section{Data calibration}\label{sec:calib}

Data calibration consists in subtracting the instrument null floor from the null depths obtained on the science object. The instrument null floor, sometimes also called the transfer function, is the response of the instrument to an unresolved object. As explained above, it is not necessarily zero because of instrumental imperfections such as phase errors, intensity mismatch, and tip-tilt variations. Because these perturbations can vary over time (e.g., phase noise decreasing with elevation), the pointings on the science object are interleaved with pointings on calibrator stars. We have adopted a total time per pointing of approximatively 20 to 25 minutes resulting from a trade-off between data acquisition efficiency and the need for measuring the null floor regularly. 

Since the calibrators are not fully unresolved, the first step to estimate the instrumental null floor is to correct the calibrator measurements for the finite extension of the stars. This correction is done using a linear limb-darkened model for the geometric stellar null \citep{Absil:2006,Absil:2011b}:\\
\begin{equation}
N_{\rm star} = \left(\frac{\pi B \theta_{\rm LD}}{4\lambda}\right)^2\left(1-\frac{7u_\lambda}{15}\right)\left(1-\frac{u_\lambda}{3}\right)^{-1};\\
\label{eq:geom}
\end{equation}
\noindent where $B$ is the interferometric baseline, $\theta_{\rm LD}$ is the limb-darkened angular diameter of the photosphere, $\lambda$ is the effective wavelength, and $u_\lambda$ is the linear limb-darkening coefficient. Given the relatively short interferometric baseline of the LBTI, this correction is generally small in the mid-infrared and the effect of limb-darkening negligible. Assuming $u_\lambda$ = 0, the typical geometric stellar null for our targets is $\sim10^{-5}$ with an error bar of $\sim10^{-6}$, which is small compared to our measurement errors (see Appendix~\ref{app:calib} for more details). Similar conclusions are obtained if we assume $u_\lambda$ = 0.5. 

After this correction, the next step in our calibration approach is to convert the null depth measurements per OB (see Figure~\ref{fig:null}, left) to a single value per pointing. Because the background bias (see Section~\ref{sec:flux}) is correlated with the nod position in a given pointing, the null depth per pointing is computed in two steps. First, the null depth for a given nod position in the current pointing $N_{p,n}$ is computed using the maximum-likelihood estimator for a Gaussian distribution (i.e., the weighted mean): 
\begin{equation}
N_{p,n} = \frac{\sum_i N_i/\sigma^2_{i}}{\sum_i 1/\sigma^2_{i}}\, ,\\
\label{eq:nullpn}
\end{equation}
where $N_i$ is the null depth of the $i$th OB and $\sigma_{i}$ the corresponding error bar (obtained from the NSC fit). The error bar on $N_{p,n}$ is computed as the propagated statistical error on the weighted mean:  
\begin{equation}
\sigma_{p,n} = \frac{1}{\sqrt{\sum_i1/\sigma^2_i}}\, ,\\
\label{eq:stat1}
\end{equation}
which decreases with the number of data points but is nonzero if all noisy measurements happen to be equal. The null depth per pointing $N_p$ is then computed as the mean value of the $N_{p,n}$: 
\begin{equation}
N_p = \frac{\sum_n N_{p,n}}{2}\, ,\\
\label{eq:nullp}
\end{equation}
where the factor 2 is the number of different nod positions (see $N_p$ for the $\beta$~Leo sequence in the right part of Figure~\ref{fig:null}). The error bar on the null depth per pointing ($\sigma_{\rm tot}$) is computed as the quadratic sum of the statistical ($\sigma_{\rm stat}$) and the systematic ($\sigma_{\rm sys}$) error terms. The statistical term is computed as 
\begin{equation}
\sigma_{\rm stat} = \frac{\sqrt{\sum_n \sigma^2_{p,n}}}{2}\, .\\
\label{eq:stat}
\end{equation}
The systematic term is composed of two different terms. The first one comes from the uncertainty on the stellar diameter ($\sigma_{\rm diam}$), which is fully correlated between all the OBs of the same pointing and hence remains the same whatever the number of OBs. The second systematic term ($\sigma_{\rm exc}$) accounts for possible measurement biases between different OBs and, in particular, the background bias. It is estimated as the square root of the excess variance \citep[e.g.,][]{Vaughan:2003}, which is the difference between the unbiased variance of the null depth measurements $S$ and that expected based on the individual error bars $\mean{\sigma_i^2}$: 
\begin{eqnarray}
\sigma_{\rm exc} &=& \sqrt{S^2-\mean{\sigma_i^2}} \, ,\\
			     &=& \sqrt{\frac{1}{n_{\rm ob}-1}\sum_i(N_i-N_p)^2-\frac{1}{n_{\rm ob}}\sum_i\sigma^2_i} \, ,
\label{eq:sys}
\end{eqnarray}
where $n_{\rm ob}$ is the number of OBs in the considered pointing. The expression of the total variance is not weighted here since systematic errors cause measurement errors that are not captured by the individual error bars. Note that this systematic error term can be underestimated if the individual error bars derived by the NSC are overestimated. Therefore, we use a systematic error floor $\sigma_{\rm flo}$ as the minimum possible value for $\sigma_{\rm exc}$. Assuming that it is dominated by the background bias, it can be estimated by looking at an empty region of the detector. Taking the average result of the whole $\beta$~Leo sequence (excluding the fourth pointing), we find a background bias of 1\,mJy per pointing. Adding the systematic error terms quadratically (valid for independent variables), the final systematic error per pointing is given by:
\begin{equation}
\sigma_{\rm sys} = \sqrt{\sigma^2_{\rm diam}+{\rm MAX}(\sigma_{\rm exc},\sigma_{\rm flo})^2}\, .\\
\label{eq:stat}
\end{equation}
Table \ref{tab:nullp} gives the null depths and error terms for the six different pointings obtained on February 8, 2015. The calibrator null depths agree very well with each other (within 0.03\%) and the total uncertainty on the null floor, represented by the dashed line in Figure~\ref{fig:null}, amounts to only 0.025\%. The two science pointings also agree relatively well (within 0.04\%) but it must be noted that the second science pointing shows a significantly larger error bar than the first one. This is due to a clear background bias between the two nod positions that appeared due to the longer mean nodding period used for this pointing caused by loop instability problems. This background bias is captured by the systematic error term and dominates the total error for pointing 4. 

\begin{table}[!t]
\begin{center}
\setlength{\tabcolsep}{4pt}
\caption{Measured null depths and corresponding 1-$\sigma$ uncertainties for the six pointings obtained on February 8, 2015 (see main text for error term definition). $F_{\rm \nu,N'}$ is the N'-band flux density computed by SED fit following the approach used in \cite{Weinberger:2015}.}\label{tab:nullp}
\begin{tabular}{c c c c c c c c}
\hline
\hline
   & Name & $F_{\rm \nu,N'}$[Jy] & $N_p$[\%]  & $\sigma_{\rm stat}$[\%] & $\sigma_{\rm sys}$[\%] & $\sigma_{\rm tot}$[\%]\\
\hline
1  & HD\,104979  & 6.1 & 0.046 & 0.047              & 0.016             & 0.050             \\
2  & $\beta$~Leo & 5.4 & 0.540 & 0.043              & 0.018             & 0.047             \\
3  & HD\,109742  & 4.2 & 0.073 & 0.051              & 0.024             & 0.056             \\
4  & $\beta$~Leo & 5.4 & 0.501 & 0.041              & 0.149             & 0.154            \\
5  & HD\,108381  & 6.2 & 0.073 & 0.050              & 0.016             & 0.053             \\
6  & HD\,109742  & 4.2 & 0.044 & 0.042              & 0.024             & 0.050             \\
\hline
\end{tabular}
\end{center}
\end{table}%

The next step in the data calibration is to estimate the instrumental null floor at the time of the science observations. It can be estimated in various ways, using for instance only bracketing calibrator measurements, a weighted combination of the calibrators, or a polynomial interpolation of all calibrator measurements. Because the calibrator stars are chosen close in magnitude and position on the sky to the science target, the instrumental null floor is generally well-behaved for the duration of the observations and we use the latter approach. A constant value is actually generally sufficient to get a good fit as shown by the solid line in Figure~\ref{fig:null}. The calibrated null is then computed as the difference between the science null depth measurement and the null floor value at the same time. The uncertainty on a calibrated null depth measurement is computed as the quadratic sum of its own uncertainty and the total uncertainty on the instrumental null floor. The latter is computed following the same approach as that used to derive the null depth uncertainty per pointing (i.e., using the quadratic sum of the statistical and the systematic uncertainties defined respectively by Equations~\ref{eq:stat1} and \ref{eq:sys}). This approach allows to estimate the systematic error on the null floor by using the distribution of calibrator null depth measurements per pointing. The final calibrated null depth measurements (or source null) for $\beta$~Leo amount to 0.478\% $\pm$ 0.050\% and 0.439\% $\pm$ 0.156\% for the first and second pointings respectively. This is consistent with the source null measured at higher spatial resolution by the KIN in the 8-13\,$\upmu$m waveband \citep[i.e., $0.42\% \pm 0.19\%$][]{Mennesson:2014}, which suggests that the disk is relatively extended. The scientific interpretation of this result will be the subject of an upcoming paper also including direct imaging L-band LBTI data of $\beta$~Leo (Hinz et al.\ in prep).  


\section{On-sky performance}\label{sec:perfo}
\subsection{Validation tests}

\begin{figure}[!t]
	\begin{center}
		\includegraphics[width=8.5 cm]{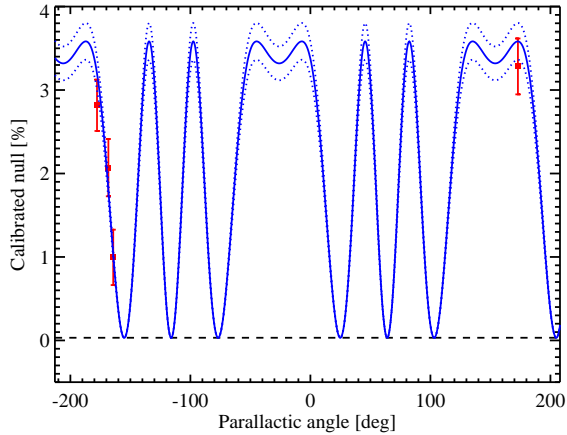}
		\caption{Calibrated nulls obtained on the bright G8III$+$A3IV binary system $\gamma$~Per (angular separation of 252\,mas) on UT December 31, 2013. The solid blue line shows the expected source null using the well-known orbital parameters for this system and the estimated flux ratio in the N' band (see main text for more information). The dotted lines correspond to the 1-$\sigma$ uncertainty on the flux ratio while the black dash line represents the geometric null floor due to the finite extension of the primary.} \label{fig:validation}
	\end{center}
\end{figure}

As part of the commissioning phase, we carried out a series of tests to examine the absolute null accuracy of the system and validate the data reduction pipeline. The first test consisted in observing a bright binary system with well-known orbital parameters and comparing the null depth measurements with theoretical predictions. We observed the double-lined spectroscopic, visual and photometric binary system $\gamma$~Per (HD~18925, G8III$+$A3V, 79\,pc) on UT December 31, 2013, using only coarse fringe tracking (see Section~\ref{sec:sensing}). We obtained one pointing on $\gamma$~Per and two bracketing pointings on calibrator stars: HD\,6860 ($\beta$\,And, M0III) and HD\,14872 (65\,And, K4III). The pointing on $\gamma$~Per consisted in four null OBs, each containing five thousand 23-ms long frames, taken at a single nod position and interleaved with background measurements. The pointings of the calibrator stars contained only one OB (or five thousand 23-ms long frames in this case). The limb-darkened angular diameter of the calibrators was obtained from the literature: 13.75 $\pm$ 0.137\,mas for HD\,6860 \citep[][]{Mozurkewich:2003} and 3.28 $\pm$ 0.056\,mas for HD\,14872 \citep[][]{Borde:2002}.

In order to estimate the expected source null, we use Equation~A5 in \cite{Mennesson:2011}. To first order, this expression relies on 5 parameters: the angular diameter of each component, their angular separation, the flux ratio at the observing wavelength, and the position angle of the secondary component. Whereas the secondary component (the A3V star) is within the diffraction limit of a single aperture ($\lambda$/D = 275\,mas), it was directly visible in our images at null and found at the predicted position \citep[i.e., an angular separation of 252\,mas and a position angle of 244~degrees using the orbital elements in][]{Pourbaix:1999}. For the flux ratio in the N' band, there are no direct measurements in the literature. The component individual visual magnitudes were estimated from the eclipse to be V$_{\rm mag}$ = 3.25 for the G star and 4.49 for the A star \citep{Griffin:1994}. Using standard color tables \citep{Ducati:2001}, the magnitudes at N band are respectively 1.50 and 4.51, which give a total magnitude of 1.43. This is approximately 0.5 magnitude fainter than the expected total magnitude measured by WISE \citep[i.e., 0.92 $\pm$ 0.06 in band 3 or 0.86 $\pm$ 0.06 in the N' band,][]{Cutri:2013}. This is not really a surprise since this system has been qualified as overluminous (with respect to model predictions) by various authors \citep[e.g.,][]{McAlister:1982,Popper:1987,Pourbaix:1999}. The same conclusion can be obtained simply by comparing the total K-band magnitude of the system \citep[i.e., 0.954 using the Johnson bright star photometry and the transformation in][]{Sierchio:2014} to the ALLWISE value of 0.92. Assuming that the primary G8III giant star accounts for this discrepancy, the individual magnitude at N band are N$_{\rm mag}$ = 0.89 $\pm$ 0.06 for the G star and N$_{\rm mag}$ = 4.51 for the A star, which corresponds to a flux ratio of 3.55\% $\pm$ 0.22\%. This is in good agreement with the best-fit flux ratio measured with our observations (i.e., 3.25\% $\pm$ 0.40\%, see Figure~\ref{fig:validation}). Note that these measurements have been obtained using only coarse fringe tracking (see Section~\ref{sec:sensing}) and without applying the correction for high-frequency phase noise (see term $\sigma_\epsilon^2$ in Equation~\ref{eq:null3}). Therefore the precision on the calibrated nulls does not represent that of the instrument at the end of the commissioning phase. 

For the second test, we observed a previously-known exozodiacal disk around a HOSTS target. These observations were reported in \cite{Defrere:2015} and confirm the detection of warm exozodiacal dust around $\eta$~Crv. Additional tests will be carried out during the science validation phase to check repeatability and high-accuracy absolute null calibration. 

\subsection{Sensitivity and throughput}

Sensitivity and throughput are two important metrics tracked regularly during the commissioning phase. The throughput is estimated both theoretically using vendor specifications for the complete LBTI optical path (3 warm reflective optics per telescope and 18 cryogenic optics in nulling mode) and experimentally using on-sky measurements of reference stars. At the end of the commissioning phase, the measured throughput was approximately 1.5 times lower than the  theoretical one (i.e., 4.5\% vs 7.5\% including the quantum efficiency of the detector of 40\%). Two culprits have been identified and will be replaced early in the science validation phase. The first one is the ZnSe uncoated window between the UBC and NIC, which was used for safety. It will be replaced by a gate valve and an automated safety system. The second one is the wavefront sensor dichroics at the entrance of the UBC which have been measured in the lab and show an absorption $\sim$10\% higher than expected. New dichroics are currently being designed. These two changes should increase the measured throughput to 7.5\%.

\begin{figure}[!t]
	\begin{center}
		\includegraphics[width=8.5 cm]{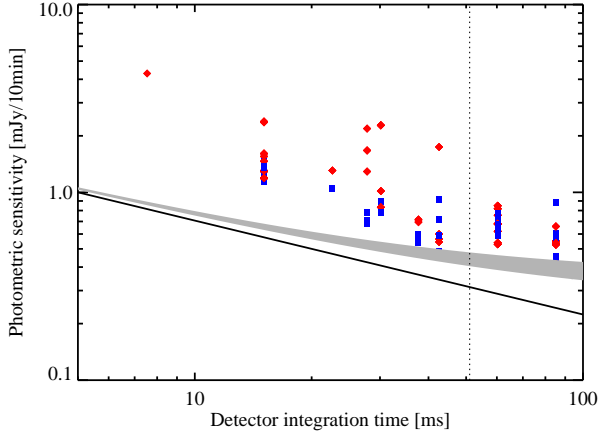}
		\caption{Measured photometric sensitivity for various representative nights of the commissioning phase (same notation as in Figure~\ref{fig:null}). The solid line shows the theoretical sensitivity for readout noise only \citep[i.e., 400 e-/pix,][]{Hoffmann:2014} while the grey-shaded area shows the sensitivity for readout noise and background noise (assuming a thermal background flux in the range of 1 to 2 Jy/pixel). The vertical dashed line indicates the integration time for which the background noise dominates the readout noise (assuming a thermal background level of 1.5Jy/pixel).}\label{fig:sensitivity}
	\end{center}
\end{figure}

\begin{figure}[!t]
	\begin{center}
		\includegraphics[width=8.5 cm]{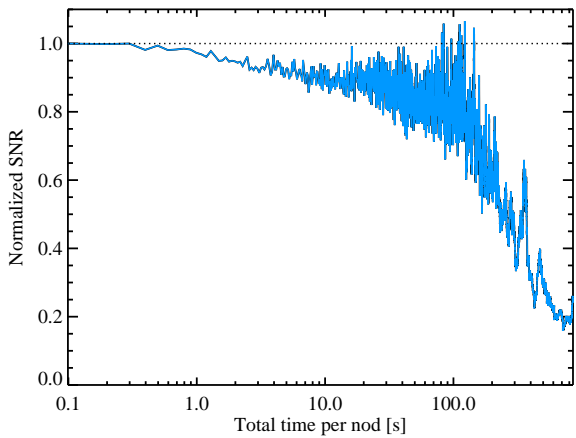}
		\caption{Normalized SNR on the background estimate as a function of the time spent per nod position (measured in the N' band in an empty region of the detector). During the acquisition of this data, the telescope was in tracking mode and went from an elevation of $56\degree$ to an elevation $41\degree$.}\label{fig:bckg_elfn}
	\end{center}
\end{figure}

\begin{figure*}[!t]
	\begin{center}
		\includegraphics[height=4.6 cm]{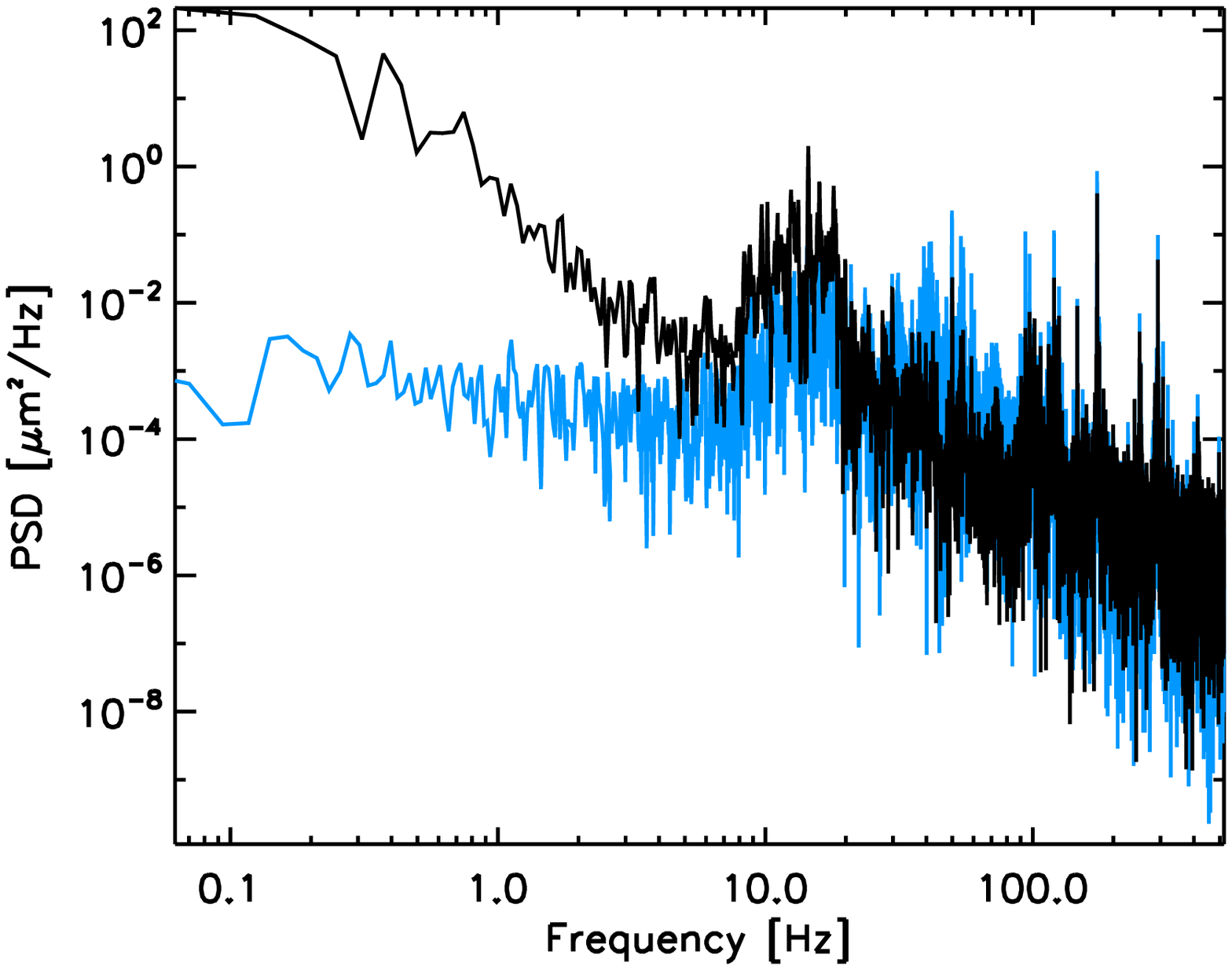}
		\includegraphics[height=4.6 cm]{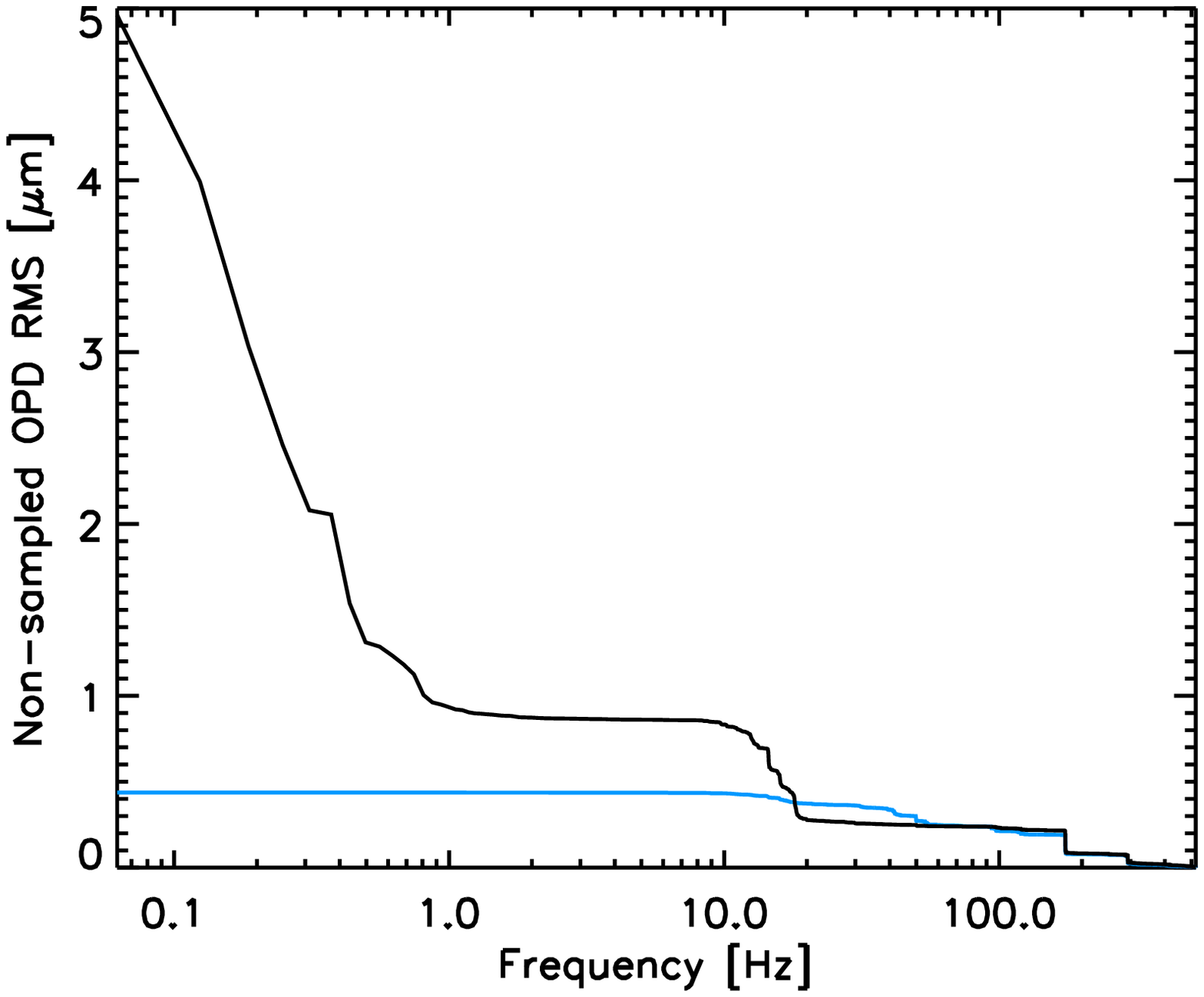}
		\includegraphics[height=4.6 cm]{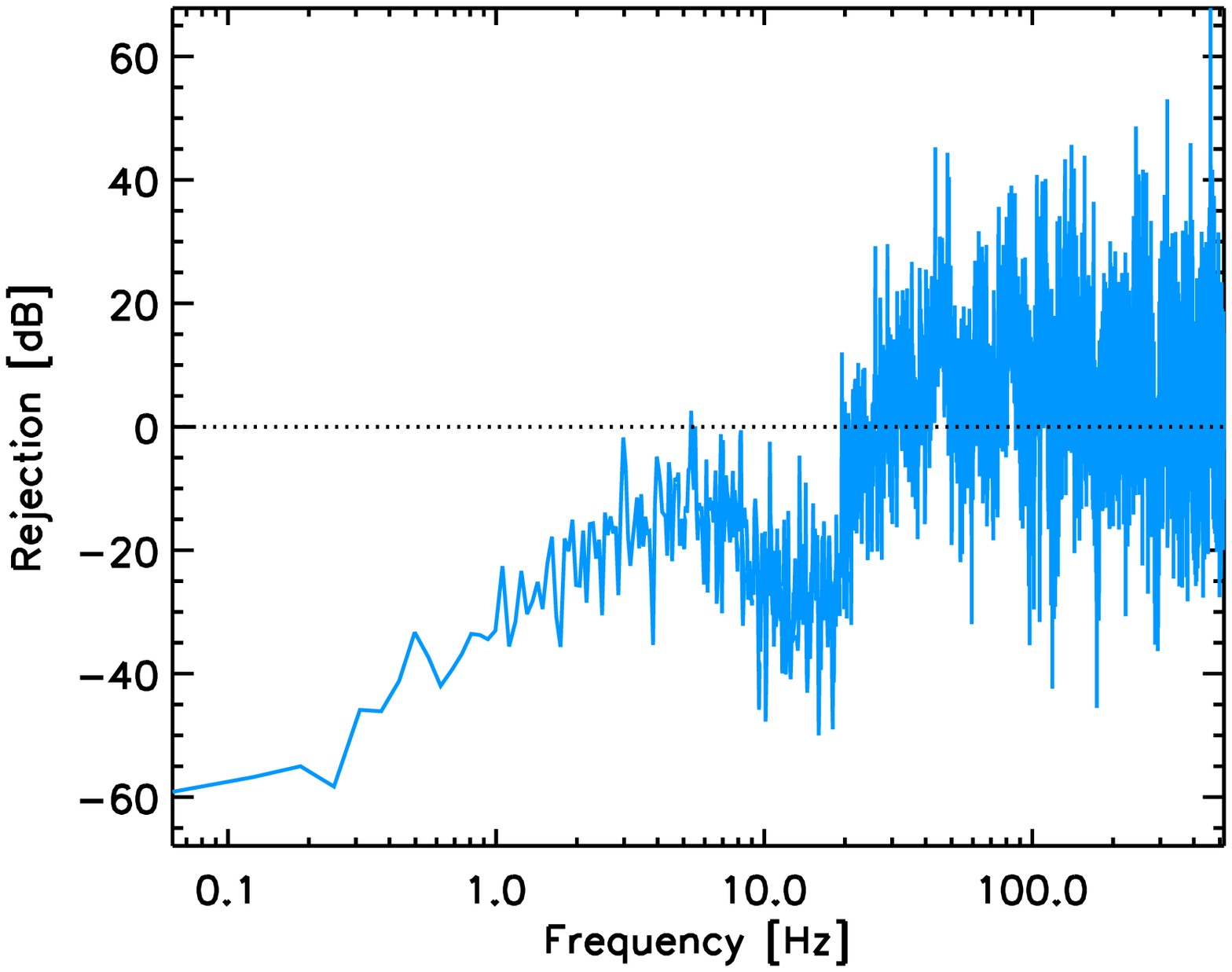}
		\caption{Left, power spectral densities of the differential OPD variations between the two AO-corrected LBT apertures in closed loop (blue line) and open loop (black line). Middle, corresponding reverse cumulative OPD variations showing the improvement in stability over 20 seconds from 5\,$\upmu$m RMS in open loop to $\sim$400\,nm RMS in closed loop. Right, corresponding frequency response. Data obtained on February 4, 2015 on the bright star $\upmu$~Gem. Loop gains were $K_p$=1, $K_i$=300, and $K_d$=0.}\label{fig:opd_perfo}
	\end{center}
\end{figure*}

The sensitivity is a critical parameter of the LBTI since it directly constrains the minimum brightness of a star that can be observed in a given time. It is limited by several factors including throughput, thermal background noise, readout noise, integration time, camera overheads, and Strehl variations. Figure~\ref{fig:sensitivity} shows the measured photometric sensitivity for various representative nights of the commissioning phase. It is computed by aperture photometry using the photometric OBs of a null sequence and scaled to coherent mode and a total integration time of 10\,min. For the maximum integration time that does not saturate the detector ($\sim$80\,ms), the photometric sensitivity amounts to 0.4-0.7\,mJy/10min, where the scatter can be explained by various factors including sky transparency, instrumental throughput variations (e.g., optical alignment), and background bias (see Section~\ref{sec:flux}). For short integration times, the photometric sensitivity is significantly worse than the theoretical sensitivity computed using the measured readout noise \citep[i.e., 400 e-/pix,][]{Hoffmann:2014} and mean measured throughput. The origin of this excess noise is related to temporal correlations induced by a combination of two effects. The first effect is inherent to the Aquarius detector, which was developed for JWST's extremely low-background operation and optimized for extremely low dark currents. This introduces an excess low-frequency noise (ELFN) in high-flux applications such as ground-based astronomy. First reported 30 years ago by \cite{Stapelbroek:1984}, this phenomenon has recently been described and characterized using the Aquarius arrays installed on VLT/VISIR \citep{Ives:2014,Kerber:2014} and LBTI/NOMIC \citep{Hoffmann:2014}. The ELFN is a form of correlated noise caused by fluctuations in the space charge induced by ionisation/recombination in the blocking layer. It manifests as a memory of photons in subsequent frames. It appears that this effect was not properly accounted for in the design of the Si:As detector material hybridized on the AQUARIUS multiplexer. 

The second contributor to the degradation of the photometric sensitivity is the background bias. Because of slowly drifting optics inside the LBTI, the spatial structure of the thermal background is not static and produces temporal variations between the background level in the photometric aperture and that estimated in the surrounding background region. This effect is illustrated in Figure~\ref{fig:bckg_elfn}, which shows the normalized SNR on the background estimate within the photometric aperture as a function of total time spent per nod position (same data as in Figure~\ref{fig:multi_bckg}). The SNR on the background estimate decreases by a factor of 2 for nods as long as $\sim$3 minutes. This is why our observing sequence is optimized to minimize the nodding period while preserving enough efficiency and enough measurements per OB for the NSC reduction. Note that this curve is only representative since the background bias depends on several factors such as the position on the detector and the elevation change rate.


\subsection{Differential phase stability}\label{sec:phase}

Differential phase variations between the two AO-corrected LBT apertures are the primary source of null fluctuations and must be controlled in real time in order to stabilize the beams out of phase. Figure~\ref{fig:opd_perfo} shows the power spectral density (PSD) of the OPD variations under typical observing conditions. It is divided in three regimes. At low frequencies ($\lesssim$10\,Hz), the OPD variations are dominated by large atmospheric perturbations of a few microns in a few seconds. At intermediate frequencies (10-50\,Hz), it is dominated by structure vibrations and in particular by a broad telescope vibration around 12\,Hz mostly due to excited eigenmodes of the swing arms that support the secondary mirrors. At high frequencies ($\gtrsim$50\,Hz), it is dominated by resonant optics located inside the LBTI cryostat, which produce large peaks at distinct frequencies such as 100, 120, and 180\,Hz. The contribution of each regime to the total OPD variation is represented by the reverse cumulative curve shown in the middle plot. Atmospheric perturbations account for a few microns in a few seconds, telescopes vibrations for 600-800\,nm between 12 and 20\,Hz, and resonant optics for 200-300\,nm at $\gtrsim$100\,Hz. In closed loop, the low-frequency component due to the atmosphere is completely removed as well as most of the telescope vibrations in the 10-20\,Hz range. Approximately 100\,nm of OPD variations are introduced by the fringe tracker in the 40-50\,Hz range due to the non-optimum tuning of the PID gains. The corresponding frequency response of the system is shown in the right-hand plot. The OPD variations are well rejected below a frequency of approximately 20\,Hz. The closed-loop residual OPD is approximately 400\,nm RMS mostly dominated by high-frequency vibrations of resonant optics inside the LBTI cryostat. Various mitigation strategies for these vibrations are currently under study. 

\begin{figure}[!t]
	\begin{center}
		\includegraphics[width=8.5 cm]{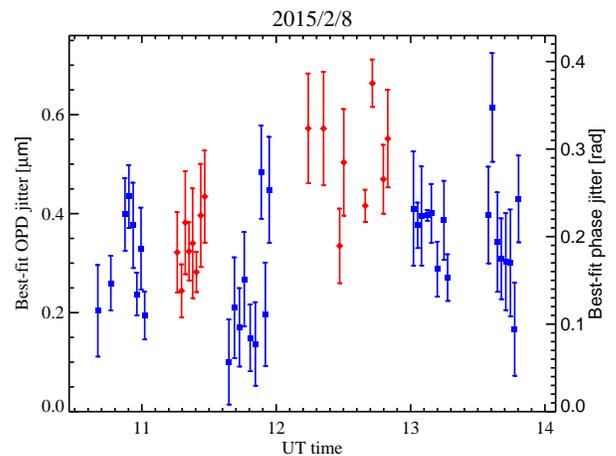}
		\caption{Best-fit phase jitter derived by the NSC as a function of UT time for the $\beta$~Leo sequence obtained on UT February 8, 2015 (PWV of 2-3\,mm). These phase variations are not captured by the near-infrared fringe tracker and, therefore, do not appear in Figure~\ref{fig:opd_perfo}.}\label{fig:nsc_phase}
	\end{center}
\end{figure}

A dominant source of phase noise not captured by Figure~\ref{fig:opd_perfo} comes from the water vapor component of the atmospheric seeing, which creates a wavelength and time-dependent phase offset between the K band, where the phase is measured and tracked, and the N' band where the null depth measurements are obtained \citep[see description in the case of the KIN in][]{Colavita:2010b}. While degrading the null stability, the existence of this term is actually crucial for accurate source null retrieval with the NSC technique since the fringe tracker completely corrects phase variations slower than the typical acquisition frequency used with NOMIC (i.e., $\lesssim$20\,Hz). As explained in Section~\ref{sec:null}, the NSC technique is inefficient to distinguish between a mean phase offset error and a true extended emission without low-frequency phase variations. For our $\beta$~Leo data obtained on UT February 8, 2015, the PWV varied between 2 and 3\,mm, which created OPD variations of 100 to 600\,nm RMS as shown in Figure~\ref{fig:nsc_phase}. 


\section{Summary and future work}

We present in this paper the data acquisition approach and data reduction method used to reach a record-setting calibrated null accuracy of 0.05\% ($1\sigma$) in the mid-infrared on the bright nearby star $\beta$~Leo. For a Sun-like star located at 10~pc, this is equivalent to an exozodiacal disk density of 15~zodi, assuming a simple ring model where the dust is confined to the habitable zone, to 30~zodi for a more physically motivated dust cloud model \citep[][]{Kennedy:2015}. Achieving this state-of-the-art contrast in the mid-infrared is the result of several key features of the LBT/LBTI such as high-quality wavefront control and low thermal background due to the adaptive secondary architecture. In particular, recent advances in co-phasing the two apertures have reduced the residual OPD jitter to 350-400\,nm RMS (at 1kHz) and considerably improved the null stability of the instrument. Another key element to get to this contrast level is the use of the NSC technique, which calibrates out important systematic errors such as mean phase setpoint variations. We present in this paper how this technique, originally developed for near-infrared nulling interferometry, has been adapted for the LBTI and modified to account for high-frequency phase variations. 

Future software work will be focused on improving the pipeline speed using a Markov chain Monte Carlo approach instead of a grid search and optimizing the observing sequence parameters for the NSC technique. We also intend to replace the least square estimator by a more general maximum likelihood function that does not rely on the assumption of Gaussian error terms. On the hardware side, several modifications have already been made to improve the optical throughput and observing efficiency. The next priority is to reduce the contributions of two dominant noise sources: the phase variations induced by the water vapor component of the atmospheric seeing and the background bias \citep[see detailed noise budget in][]{Defrere:2015b}. By reducing the former by a factor of two and solving the latter, the instrument could achieve a calibrated null accuracy of 0.01\% over a three-hour observing sequence. 



\begin{acknowledgements}
The authors are grateful to M.~Colavita for helpful discussions on fringe sensing and nulling data reduction. The Large Binocular Telescope Interferometer is funded by the National Aeronautics and Space Administration as part of its Exoplanet Exploration Program. The LBT is an international collaboration among institutions in the United States, Italy and Germany. LBT Corporation partners are: The University of Arizona on behalf of the Arizona university system; Instituto Nazionale di Astrofisica, Italy; LBT Beteiligungsgesellschaft, Germany, representing the Max-Planck Society, the Astrophysical Institute Potsdam, and Heidelberg University; The Ohio State University, and The Research Corporation, on behalf of The University of Notre Dame, University of Minnesota and University of Virginia. This publication makes use of data products from the Wide-field Infrared Survey Explorer, which is a joint project of the University of California, Los Angeles, and the Jet Propulsion Laboratory/California Institute of Technology, funded by the National Aeronautics and Space Administration. MW and GK acknowledge the support of the European Union through ERC grant number 279973. 
\end{acknowledgements}

\appendix
\section{A. Choice of calibrator stars} \label{app:calib}

\begin{figure}[!t]
	\begin{center}
		\includegraphics[height=5.4 cm]{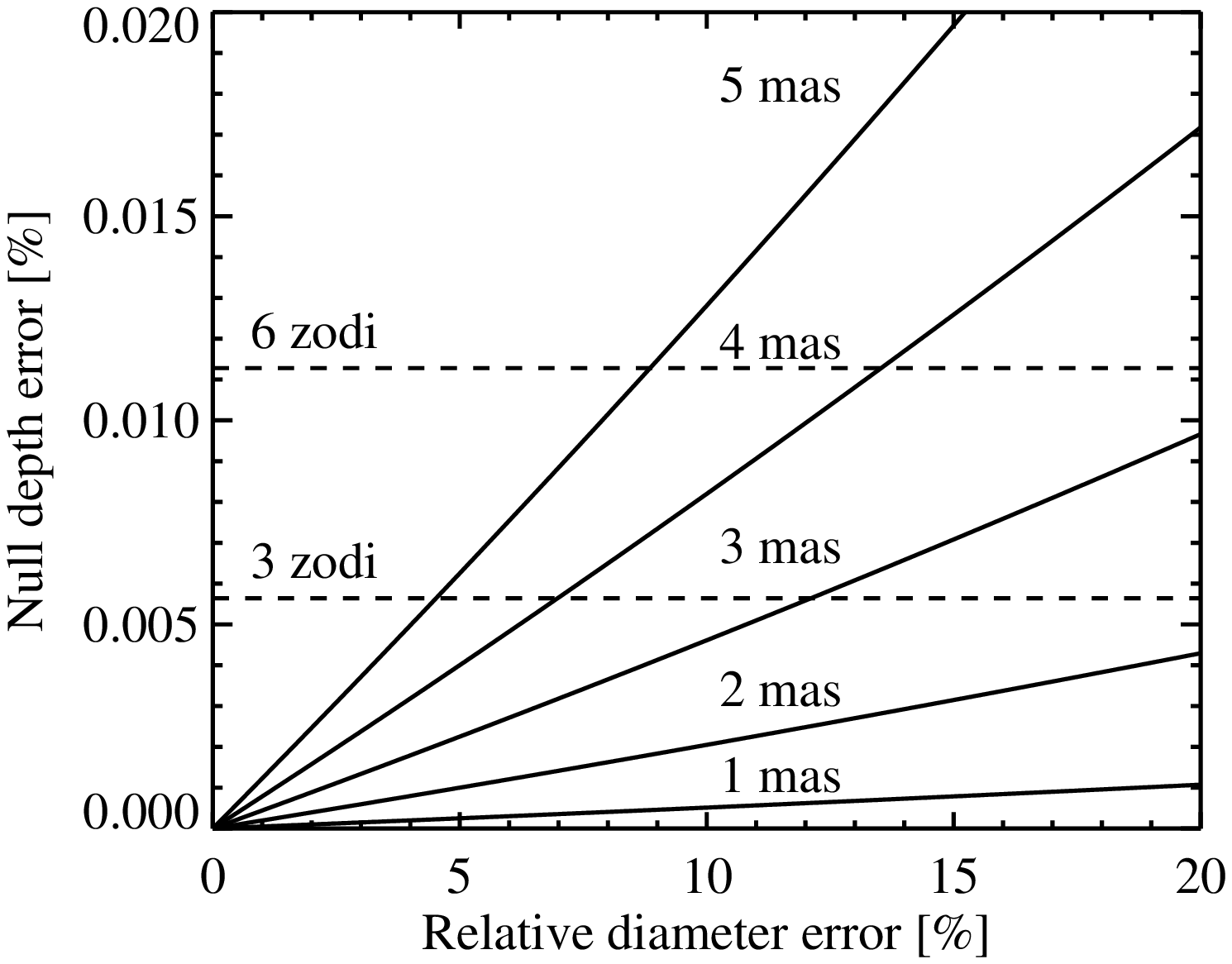}
		\includegraphics[height=5.4 cm]{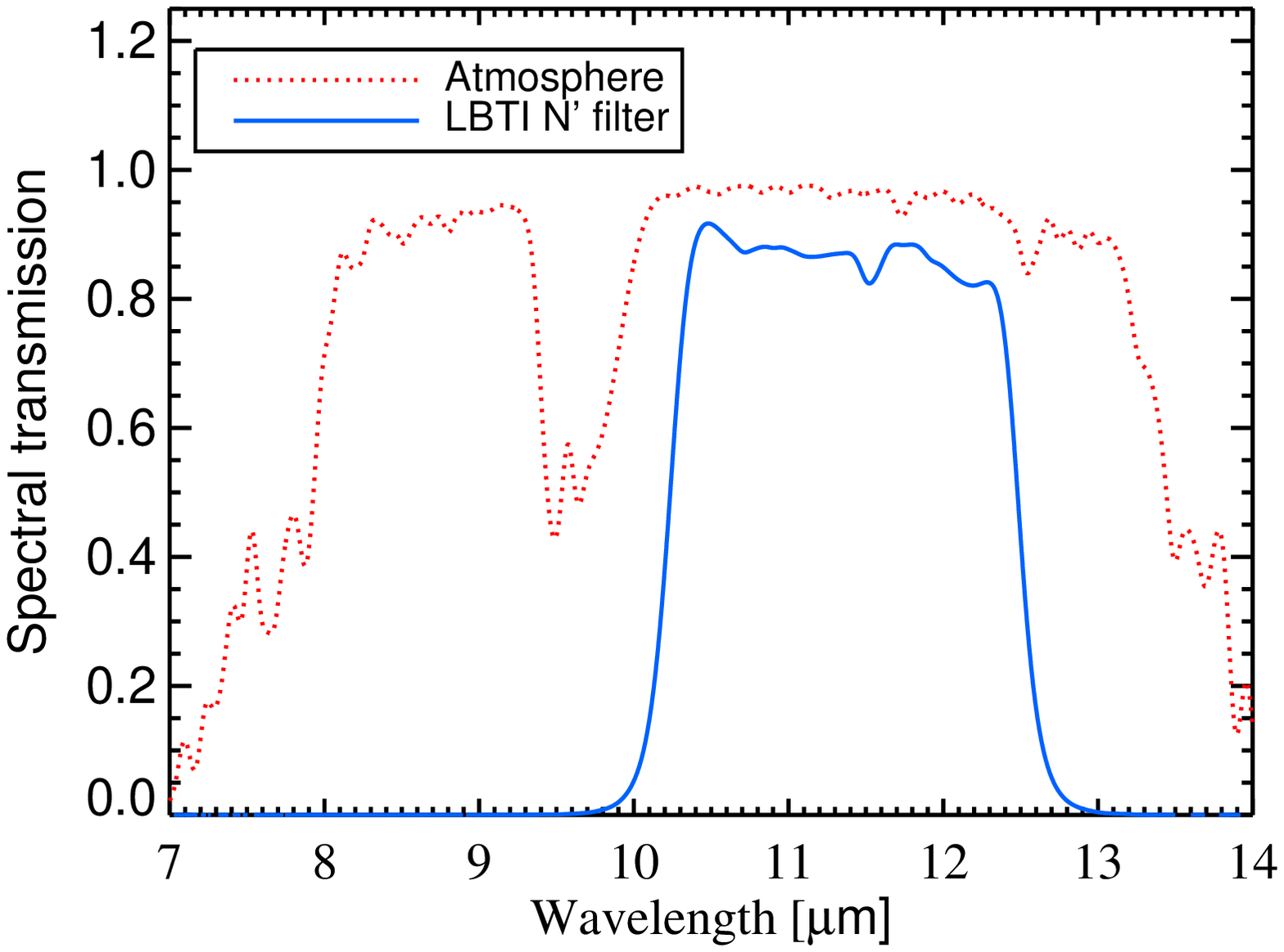}
		\caption{Left, impact of calibrator diameter uncertainty on the LBTI null uncertainty (computed at a wavelength of 11.1\,$\upmu$m). The 5 solid lines show the results obtained for calibrators of various diameters (1 to 5\,mas). The dashed lines indicate for comparison the null excesses expected for a Sun located at 10\,pc and surrounded by a face-on disk with two different zodi levels \citep[i.e., 3 and 6~zodi, defined as in][]{Kennedy:2015}. Assuming that the tolerable error due to the calibrator uncertainty has to be kept  below 3~zodi, this corresponds to a $\sim12\%$ relative error on the diameter of a 3\,mas star, but only a $\sim4\%$ relative error on the diameter of a 5\,mas star. Right, spectral transmission curve of the N' filter used to obtain the data presented in this paper (solid blue line). The dotted red line represents the infrared transmission spectrum of the atmosphere, computed for a representative observing site (i.e., Mauna Kea) and assuming 3\,mm of PWV and an air mass of 1 \citep[][]{Lord:1992}.}\label{fig:cal-err}
	\end{center}
\end{figure}

Frequent null depth measurements on calibrator stars are of primary importance in order to accurately estimate the instrumental floor and, hence, the source null of the science objects. The choice of a particular calibrator star is driven by the need for getting observations in conditions as close as possible as those at the time of the observation of the science object. Several parameters are generally considered such as the N'-band brightness, which has to be similar or slightly larger than the science target, the pointing direction, which has to be ideally within 10 degrees, and at a similar elevation, and the absence of known companion within 5$''$. The spectral type is generally driven by the need for matching the N'-band magnitude, which implies that typical calibrators are K/M giants. Several different calibrators are generally used in order to minimize the risk of selecting a bad one. If necessary, different density filters are used to match the flux count in the visible (AO system) and the near-infrared (PHASECam). 

Another important parameter is the angular size of the calibrator and the corresponding uncertainty. Figure~\ref{fig:cal-err} shows the impact of calibrator diameter uncertainty on the null uncertainty for various angular diameters. The larger the stars, the larger the resulting uncertainty in the instrumental null floor for a given uncertainty in the calibrator angular size. Typical calibrators for the HOSTS target sample have an angular diameter smaller than 2-3\,mas, which allows a relative diameter uncertainty of up to 20\% in order to keep the corresponding null uncertainty below 10$^{-4}$. Such accuracy can easily be obtained with surface-brightness relations \citep[e.g.,][]{Kervella:2004} used in other interferometric survey \cite[e.g.,][]{Absil:2013,Ertel:2014,Mennesson:2014}.

Finally, the least-demanding constraint is to choose calibrator stars sufficiently unresolved in order to produce fringes of sufficient quality in the near-infrared. Experience shows that the fringe tracker works well until an angular diameter of approximately 15\,mas, which corresponds to an absolute visibility of 0.75. 

\section{B. Transmission profile} \label{app:trans}

The right panel of Figure~\ref{fig:cal-err} shows the spectral transmission of the N' filter used to obtain the data presented in this paper (solid blue line). Its effective wavelength is 11.1\,$\upmu$m, which is well-suited to search for warm exozodiacal dust disks (according to Wien's law, the emission from a 300\,K blackbody peaks at around 10\,$\upmu$m), and its full width at half maximum amounts to 2.27\,$\upmu$m. A broadband filter centered around 8.7\,$\upmu$m and various narrowband filters are also available for more specific science observations \citep[see list in][]{Defrere:2015b}. The dotted red line represents the infrared transmission spectrum of the atmosphere, computed for a representative observing site (i.e., Mauna Kea) and assuming 3\,mm of PWV and an air mass of 1 \citep[][]{Lord:1992}.

\section{C. Impact of background region} \label{app:bckg}

\begin{figure}[!t]
	\begin{center}
		\includegraphics[height=5.4 cm]{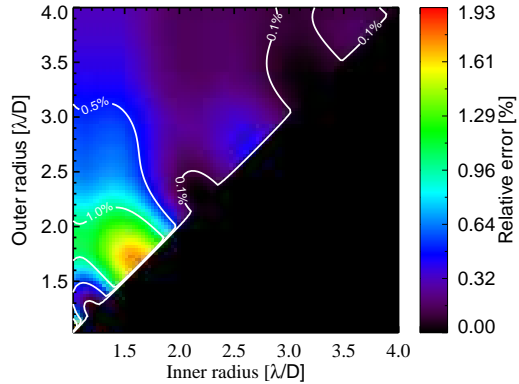}
		\caption{Relative error on the flux estimate as function of the inner and outer radii of the background annulus used for aperture photometry. The resulting error on the source null can simply be obtained by multiplying these values by the null depth. For instance, for a thin annulus located at $\sim1.7\lambda/D$ (i.e., the position of the first Airy ring) and a typical null depth of 1\%, the error on the source null amounts to $\sim$0.02\%. In practice, the position of the background annulus is chosen so that the resulting error on the source null is smaller than 0.01\%.}\label{fig:bbias}
	\end{center}
\end{figure}

As discussed in Section~\ref{sec:flux}, aperture photometry uses a circular background region located close to the center of the beam in order to estimate the background level within the photometric aperture. Assuming for the moment that the background is uniform, it is necessary to make sure that the extension of the Airy pattern into the background region does not significantly bias the background estimate (and hence the flux measurement). This effect is represented in Figure~\ref{fig:bbias}, which shows the relative error on the total flux of the star with respect to the size of the inner and outer radii of the background annulus. For an inner radius larger than $\lambda/D$, the resulting error ranges between 0 and $\sim$2\% of the measured flux (depending on the position of the background annulus). This means that, for a typical null depth of 1\%, the resulting error on the source null can be as high as 0.02\%, which is significant compared to our error bars. In practice, this bias is completely calibrated out if the science targets and the calibrators have the same magnitude and the instrumental null is stable. However, since it is not always the case, we choose the inner and outer radii of the background annulus so that the resulting error on the source null is smaller than 0.01\%. 

Note that, in practice, test data indicate that the detector shows vertical (in the altitude direction) flux correlation so that using a background annulus creates a small bias on the background estimate in the photometric aperture. Therefore, we do not always use an annulus centered on the photometric region as in classical aperture photometry but two regions that cover the same columns as those used in the photometric aperture \cite[this is the technique used in][]{Defrere:2015}. The two background regions are arranged symmetrically around the photometric region to remove any first-order slope in the background. The sigma-clipped average of each column is then computed and subtracted from each pixel in the corresponding column of the photometric aperture. However, since the background shows additional irregularities, even after nod subtraction (see description of background bias in Section~\ref{sec:flux}), this strategy does not always give the best results. 

\bibliographystyle{apj}

\end{document}